%% file: 000_Puppeteering.tex
\documentclass[sigconf,anonymous=false]{acmart} 
\renewcommand\footnotetextcopyrightpermission[1]{} 
\usepackage{todonotes}
\usepackage[
    nolist
    ]{acronym} 
\usepackage{listings}
\usepackage{multirow}
\usepackage{subfigure}
\usepackage{float}
\usepackage[ruled,linesnumbered,noend]{algorithm2e}

\newcommand{\BS}{P_{\mathrm{bootstrap}}}
\newcommand{\SW}{P_{\mathrm{socket}}}
\newcommand{\PS}{P_{\mathrm{shared}}}
\newcommand{\LP}{P_{\mathrm{local}}}
\newcommand{\EP}{P_{\mathrm{extracted}}}
\newcommand{\TT}{T_{\mathrm{trace}}}
\newcommand{\TC}{T_{\mathrm{crawl}}}

\renewcommand{\setminus}{-}

\AtBeginDocument{%
  \providecommand\BibTeX{{%
    \normalfont B\kern-0.5em{\scshape i\kern-0.25em b}\kern-0.8em\TeX}}}

\begin{document}\pagestyle{plain} 

\input{acronyms}

\title{Malware Sight-Seeing: Accelerating Reverse-Engineering via Point-of-Interest-Beacons}

\author{August See}
\authornote{Both authors contributed equally to this research.}
\email{see@informatik.uni-hamburg.de}
\affiliation{
  \department{IT-Security and Security Management}
  \institution{University of Hamburg}
  \city{Hamburg}
  \country{Germany}
}

\author{Maximilian Gehring}
\authornotemark[1]
\email{maximilianheinrich.gehring@stud.tu-darmstadt.de}
\affiliation{
  \department{Telecooperation Lab}
  \institution{Technical University of Darmstadt}
  \city{Darmstadt}
  \country{Germany}
}

\author{Max Mühlhäuser}
\email{max@tk.tu-darmstadt.de}
\affiliation{
  \department{Telecooperation Lab}
  \institution{Technical University of Darmstadt}
  \city{Darmstadt}
  \country{Germany}
}

\author{Mathias Fischer}
\email{mfischer@informatik.uni-hamburg.de}
\affiliation{
  \department{IT-Security and Security Management}
  \institution{University of Hamburg}
  \city{Hamburg}
  \country{Germany}
}

\author{Shankar Karuppayah}
\email{kshankar@usm.my}
\affiliation{
  \institution{National Advanced IPv6 Centre}
  \institution{Universiti Sains Malaysia}
  \city{Penang}
  \country{Malaysia}
}

\begin{abstract}
New types of malware are emerging at concerning rates. However, analyzing malware via reverse engineering is still a time-consuming and mostly manual task. For this reason, it is necessary to develop techniques that automate parts of the reverse engineering process and that can evade the built-in countermeasures of modern malware. 
The main contribution of this paper is a novel method to automatically find so-called \textit{\acp{POI}} in executed programs. \acp{POI} are instructions that interact with data that is known to an analyst. They can be used as beacons in the analysis of malware and can help to guide the analyst to the interesting parts of the malware. Furthermore, we propose a metric for \acp{POI}, the so-called confidence score that estimates how exclusively a \ac{POI} will process data relevant to the malware. 

With the goal of automatically extract peers in P2P botnet malware, we demonstrate and evaluate our approach by applying it on four botnets (\textit{ZeroAccess}, \textit{Sality}, \textit{Nugache}, and \textit{Kelihos}).
We looked into the identified \acp{POI} for known IPs and ports and, by using this information, leverage it to successfully monitor the botnets. 

Furthermore, using our scoring system, we show that we can extract peers for each botnet with high accuracy. 

\end{abstract}


\maketitle

\input{010_introduction}
\input{020_background}
\input{030_pointsofinterest}

\input{040_pinpuppet}
\input{060_evaluation}
\input{070_relatedwork}
\section{Conclusion}
\label{sec:conclusion}
The reverse-engineering of malware is a crucial and labor-intensive task. In this paper, we introduce a new method to guide reverse-engineers in analyzing malware via \aclu{POI} beacons. 
These \acp{POI} are instructions that process data in which an analyst is interested. 
We introduce a method to identify them automatically and rate them based on how often a \ac{POI} is used by the binary to access relevant data.

We demonstrate the usage of our method by applying it to monitor P2P botnets to show that \acp{POI} can be used for advanced, malware analysis without prior reverse engineering.. 
Using our PinPuppet prototype, we demonstrate how both the identification and the scoring of \acp{POI} works. 
Our evaluation indicates that we could successfully leverage PinPuppet to crawl four different botnets.  
Utilizing our scoring mechanism we were able to significantly reduce the number of falsely extracted peers in our crawling primitive.

As future work, we would like to further investigate on  automating our PinPuppet by integrating it to a sandbox, so that all required information such as IPs are obtained automatically. 
In addition, we would also like to explore extending our PinPuppet and \ac{POI} approaches on botnets with destination-dependent payloads.

\bibliographystyle{ACM-Reference-Format}
\bibliography{puppet}

\input{999_appendix}

\end{document}

%% file: acronyms.tex
\begin{acronym}
    \acro{POI}{Point-of-Interest}
    \acro{MM}{membership maintenance}
    \acro{C2}{Command and Control}
    \acro{gcc}{GNU Compiler Collection}
    \acro{MSB}{most significant byte}
    \acro{LSB}{least significant byte}
    \acro{DBI}{dynamic binary instrumentation}
    \acro{RPC}{Remote Procedure Call}
    \acro{LICA}{Less Invasive Crawling Algorithm}
    \acro{BFS}{Breadth first search}
    \acro{DFS}{Depth first search}
    \acro{ASLR}{Address Space Layout Randomization}
    \acro{P2P}{Peer-to-Peer}
    \acro{JSON}{JavaScript Object Notation}
    \acro{OS}{operating system}    
    \acro{VM}{virtual machine}
\end{acronym}

%% file: 010_introduction.tex
\section{Introduction}

Malicious software (malware) is a severe threat to every IT system and its numbers are steadily increasing. In 2018 alone, Symantec reported \cite{Symantec2019}  more than $240$ Million new variants of malware. 
Security researchers and analysts are always in the look out for new emerging malware as well as existing ones to protect end users and organizations. 
To understand a new type of malware, analysts often need to reverse engineer it by focusing on the inner-workings of the malware.
This process helps to derive unique signatures of its behavior, which can be used in security tools like antivirus software and intrusion detection systems. 

Reverse-engineering is mostly a manual task and requires a specialized and trained personel. 
Thus, the automation of reverse-engineering to support the human analyst is an ongoing research field and has led to an ongoing arms race between analysts and malware authors. 
Meanwhile, more advanced reverse engineering techniques encompass methods for the replay of protocols using symbolic execution \cite{replayer}, analysis of data structures \cite{autoRe}, taint tracking of data \cite{dynamictaintDytan}, and sophisticated reverse-engineering suites and tools like IDA\footnote{https://www.hex-rays.com/products/ida/} and Ghidra\footnote{https://ghidra-sre.org/}. 
Nevertheless, even these advanced techniques fall short in the analysis of novel malware equipped with advanced countermeasures to impede its analysis \cite{antiTaint07,antiTaint13,symbolicExecObfuscation}.

In this paper, we propose a new reverse-engineering technique to identify so-called \acp{POI} in applications that can guide analysts as beacons in the process of analysing malware. 
Our method exploits that all data used by an application is at some point is either loaded in a register, written to, or read from the memory.
\acp{POI} are instructions that interact with data that is known to an analyst. 
Thus, using \textit{a priori} information, e.g., \ac{C2} IP addresses, our proposed technique leverages dynamic binary instrumentation tools to identify specific instruction addresses within a malware binary that are used to access these values during their execution. 
This may result in a large number of \acp{POI} from general purpose functions such as memcpy or strcpy. 
The fact that the searched data is copied using a general purpose function may not necessarily be important information for the analyst. 
However, the points at which the searched data is mainly processed are of interest. 
A so-called confidence score is therefore calculated for the \acp{POI}. 
This score indicates how exclusively a \ac{POI} handles the searched data.
Based on the scores, analysts can focus on \acp{POI} that are useful to them. 

As a proof of concept, we demonstrate the feasibility of using \acp{POI} to automatically monitor four \ac{P2P} botnets without the need to reverse-engineer them. 
This involves extracting IPs and ports from a running bot. 
In evaluating our proof of concept implementation, we show that the confidence score overestimates the quality for only $19$ out of a total of $400$ \acp{POI}. 
It is thus a good conservative estimator for the quality of the results that is expected. 
For all botnets we analyzed, our filtering based on the confidence score improves the results.
For instance, within the ZeroAccess botnet, the number of incorrectly extracted peers drops from $59.98$ to zero while retaining correct results.

The remainder of this paper is structured as follows. In \autoref{sec:background}, we introduce the necessary background in reverse engineering and botnet malware. \autoref{sec:poi} introduces our proposed technique and \autoref{sec:pinpuppet} details the implementation of our \ac{POI} technique and the setup for automatic monitoring of P2P botnets. Then, \autoref{sec:evaluation} discusses our evaluation along with the corresponding results. 
Finally, \autoref{sec:rel-work} introduces the related work and \autoref{sec:conclusion} concludes this paper.

%% file: 020_background.tex
\section{Background}
\label{sec:background}
This section provides an overview of reverse engineering techniques, anti-reverse engineering techniques employed by malware authors, P2P botnets, and botnet monitoring.

\subsection{Reverse Engineering}
\label{sec:bg_rev_eng}

Reverse engineering can be divided into static and dynamic analysis. 
Both methods aim to reconstruct and modify the functions, objectives, or artifacts of a program.
For this purpose, binaries are often analyzed at assembler instruction level. When manually performing this process, reverse engineers often times rely on common patterns or contextual information like annotations, used strings or function names to infer the purpose of a code region. This information behaves as beacons \cite{votipkaObservationalInvestigationReverse2019a} and can accelerate the work because the reverse engineer already has clues where to start his work, so that it is not necessary to start at the entry point of the binary. Tools like sandboxes or software reverse engineering suites can provide further static and dynamic information about a malware sample. The information collected by such tools ranges from created files, contacted IPs to marking instructions with certain functionality in the binary\footnote{\url{https://www.hex-rays.com/products/ida/lumina/}}.
In most cases, both methods are used to complement each other.

\textit{Static analysis} is everything that can be obtained about the program without executing it. 
This include the knowledge about the functionality of the binary, as well as the extraction of data such as strings and metadata.
Well known tools for this process are e.g., IDA and Ghidra.
Both tools bring functionalities like identifying individual functions, finding strings, and locating where system functions are called.

In contrast to static analysis, \textit{dynamic analysis} executes actual parts of the program in an analysis environment.
This allows runtime information to be extracted from the binary. A very common type of tool used for dynamic analysis is a debugger like GDB\footnote{https://www.gnu.org/software/gdb/}.
\textit{\Ac{DBI}} is a kind of dynamic analysis. With \ac{DBI}, new code is inserted into a binary at runtime. This code can access registers and analyze the data that instructions are reading and writing. This way one can log instructions and accessed data, i.e., registers and memory, for every executed instruction. The code can be added at different granularity's, e.g., for every instruction or for every function.
Furthermore \ac{DBI} can also be used to change an application's behavior, e.g., by modifying registers. \Ac{DBI} is often faster than conventional debuggers, since no breakpoints have to be set which generate an interrupt and therefore a context switch to the debugger~\cite{eilam2011reversing}.
Frameworks in this area are among others Intel Pin~\cite{intelpin} and DynamoRIO~\cite{dynamorio}.

As malware authors want to prevent reverse engineering, they develop more and more sophisticated anti-reverse engineering techniques. These are meant to make the work of the reverse engineer more difficult. A detailed overview is given by the authors of \cite{surveyMalware,yan2008revealing,malwareObfuscation}, however, in the following we will highlight two such techniques.

Packers change a binary so that (a part) of the binary is encrypted. This part is then decrypted when the binary is executed. A packer may not be deterministic, it can decrypt the same malware multiple times and creates every time a binary with a different hash value~\cite{yan2008revealing}.
    
Debugger detection techniques prevent or detect the presence of a debugger (e.g. \textit{checkRemoteDebuggerPresent\footnote{https://docs.microsoft.com/en-us/windows/win32/api/debugapi/nf-debugapi-checkremotedebuggerpresent}, isDebbugerPresent\footnote{https://docs.microsoft.com/en-us/windows/win32/api/debugapi/nf-debugapi-isdebuggerpresent})}. This can be a simple search for processes or window names up to the detection of artifacts in memory or in the execution environment~\cite{antiDebugging}.

\subsection{P2P Botnets and Botnet Monitoring}
\label{sec:bg_p2p_botnets}
Botnets can be divided into different categories based on the network architecture they use.
Depending on the architecture, a botnet can, for example, be deployed easier or harder, be more resilient to takedowns, or be more efficient in internal communication \cite{karuppayah2018advanced}.
P2P botnets do not use dedicated \ac{C2} servers, but are, as the name suggests, connected in a peer-to-peer network.

Each bot in a P2P botnet has a peer list that contains the information necessary to contact other bots, such as IP addresses, ports, and unique identifiers.
In this paper we denote bootstrap list $\BS$ to the initial peer list a bot carries before it is executed.
The bootstrap list is used by a bot after infecting a machine to join the botnet.
The peer list denotes the list of peers that a bot holds after it has joined the P2P overlay of the botnet.
The peer list is not static, as new bots join the network (new infections) and other bots leave the network (removals of bots).
To ensure persistent connectivity, P2P botnets use an \ac{MM}-mechanism to maintain the peer list.
In general, this is done by periodically contacting existing peers from the peer list and, if there are too few responsive peers, requesting new ones.
Especially the message to request new peers is interesting and is called \textit{getL} message in the following.
The response to a \textit{getL} message is a \textit{retL} message containing a list of shared peers.

In P2P botnet monitoring the \textit{getL} message is used to recursively crawl for peers. This can reveal the extent of a botnet.
For this, however, the protocol that the bots speak among themselves must be known, e.g., message format and encryption. This knowledge is achieved by reverse-engineering the binary.
This requires time and expertise and at the same time the result is only valid for only one botnet. In the worst case, even only for one version of the botnet.

%% file: 030_pointsofinterest.tex
\begin{figure*}
    \centering
    \includegraphics[width=\linewidth]{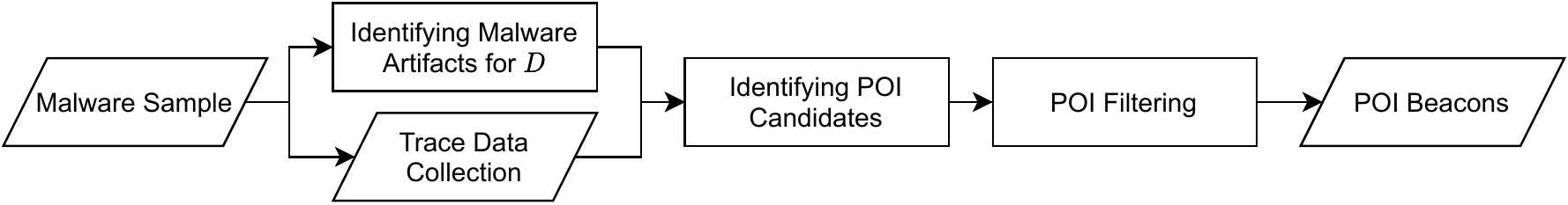}
    \caption{The \ac{POI} discovery process.}
    \label{fig:poi_discovery}
\end{figure*}
\section{POI as Beacons}
\label{sec:poi}
In this section, we first introduce the concept of Points-of-Interest (POI) as a type of beacon to accelerate reverse-engineering efforts. 
Next, we explain the steps of identifying all \ac{POI} candidates.   
Then, we discuss strategies to evaluate and pick the best \acp{POI} to be used as beacons.

\label{sec:poi_definition}

As discussed in \autoref{sec:background}, beacons can serve as an excellent guidance for malware analysts to focus and accelerate their reverse engineering efforts.  
Beacons are essentially patterns and markers that are used by analysts to either focus or ignore certain parts of the binary, e.g., known encryption routines or anti-reverse engineering components \cite{votipkaObservationalInvestigationReverse2019a}.

As most malware are programmed to carry out some read/write operations on the infected machine, e.g., registry, file,  or network access, this behaviour would also be reflected at the instructions level of the binary of the malware sample.
During the execution of these instructions, various data would be accessed through registers and memory regions of the infected machine which are often managed directly by the \ac{OS}.
If the data/value accessed by the malware is known, e.g., specific file names or IP addresses, one could observe the data being accessed by specific instruction address(es) within the binary, i.e., either through memory or registers.
These addresses are referred by us as \acp{POI}. 
In reality, the details of the exact data accessed by the malware can be easily obtained through behavioural analysis carried out using malware sandboxes.

As an example, a typical operation a malware would perform is constructing a data buffer and writing it into a configuration file. 
\autoref{lst:write_data_example} illustrates an example assembly code that represents the behavior of a malware that writes a static header to a file. 
In Lines $1-4$, the static header for the configuration is generated. 
In Lines $6-8$, the \textit{write\_config} function is called with the generated buffer data. 
Assuming the header data is known upfront, the instruction in Line 4 would represent a \ac{POI}, i.e., the instruction when the expected header data is observed in the memory.

\begin{lstlisting}[language={[x86masm]Assembler},caption={Generated Assembly Code},label={lst:write_data_example}]
mov	BYTE PTR -140[ebp], 0xfe
mov	BYTE PTR -139[ebp], 0xdc
mov	BYTE PTR -138[ebp], 0xba
mov	BYTE PTR -137[ebp], 0x98
// more configuration file generation
lea	eax, -140[ebp]
push	eax
call	write_config
\end{lstlisting}

If an analyst is interested in finding out the operations that are relevant prior or after the creation of the configuration file, the \ac{POI} can be used as a starting point of starting the analysis.
However, the process of identifying \ac{POI} candidates is not trivial and may sometime yield inaccurate results.
In the next sections, we discuss our methodology in identifying and selecting high-quality \acp{POI}.

\subsection{Methodology}
\label{sec:poi_methodology}
In this section, we briefly introduce the various stages within our methodology on extracting \ac{POI} beacons as illustrated in \autoref{fig:poi_discovery}.
We will elaborate the relevant stages in the subsequent subsections.

Firstly, a valid malware sample needs to be identified and retrieved for further analysis.
Using this sample, data or values that are accessed by the malware, e.g., IP addresses, accessed files, etc., need to be identified. 
Next, the execution trace of the malware needs to be recorded.
This can be achieved using \ac{DBI} or debuggers such as GDB (see \autoref{sec:bg_rev_eng}).

Using information obtained from the previous steps, we perform our \ac{POI} candidates identification. 
Finally, we filter out \acp{POI} to ensure we retain only high-quality \acp{POI} and use them as \ac{POI} beacons for guiding malware analysts to reverse engineer the malware.
The remainder part of this section elaborates each of the steps introduced within this methodology accordingly.

\subsection{Identifying Malware Artifacts }
\label{sec:poi_data_set}
As our methodology requires a priori information $D$ pertaining the malware sample, some information, $d \in D$, needs to be extracted from the sample.
This information can be obtained using malware sandboxes to identify artifacts introduced/accessed by the malware sample.
Alternatively, one could leverage general purpose interactions of the malware with the \ac{OS}, e.g., writing files, using the \ac{OS} API monitoring tools such as \textit{API Monitor}\footnote{\url{http://www.rohitab.com/apimonitor}}, \textit{Process Monitor}\footnote{\url{https://docs.microsoft.com/en-us/sysinternals/downloads/procmon}}, and \textit{strace}\footnote{\url{https://linux.die.net/man/1/strace}} during runtime.
Finally, the analyst could also add manual entries, e.g., expert information, to the overall set $D$.

Choosing the information for a sample is closely related to the main goal of the analyst in obtaining relevant beacons.
For instance, if the analyst is interested in studying the network communication modules of the malware, the set of IP addresses and ports that are used by the malware for network communication would be a suitable choice.
Nevertheless, it is important to ensure that only unique information are picked to increase the likelihood of generating high-quality \ac{POI} beacons.

For instance, assuming a sandbox analysis reported an outgoing connection from the malware to a remote host with the IP address $10.20.30.40$ and port $80$, it would be suitable to include only the IP address in the set $D$.
The reason for leaving the port value out would be because the (decimal) value for port $80$ could be used by other components also for network communication purposes.
Moreover, the value could also be used as an internal counter or as the ASCII character \texttt{'P'} withing the malware.
In contrast, the likelihood of observing that the data corresponding to the IP is processed within another context is rather low.

Another consideration that is worth noting would be the representation of the data added in $D$.
The data could be processed differently during the execution of a malware depending on various factors such as the computer architecture and the \ac{OS} system.
For instance, the IP address from the previous example could be accessed in any one of the following manners during execution:
\begin{enumerate}
    \item \texttt{0x0A141E28}, i.e., the IP in binary form with \ac{MSB} first.
    \item \texttt{0x281E140A}, i.e., the IP in binary form with \ac{LSB} first.
    \item The ASCII strings \texttt{"10.20.30.40"} and \texttt{"0A141E28"} if the malware handles the IP address as ASCII text. 
\end{enumerate}
Hence, whenever possible, all possible representations of the data that could uniquely identify the malware-related artifact should be included in $D$.

\subsection{Identifying POI Candidates}
\label{sec:poi_identify}

As explained earlier, \acp{POI} are defined as instruction addresses that accesses, i.e, read or write operations, registers or memory regions that has the value of any one of the elements within $D$. 
Hence, in this step, suitable \acp{POI} candidates are identified by iterating the  malware execution trace while looking for instructions that access any of the expected malware artifacts in $D$.

Nevertheless, the process of finding matching instructions is not trivial.  
The main reason for this is the manner data or values processed by an \ac{OS}. 
Depending on the CPU architecture supported by the \ac{OS}, i.e., $32$-bit vs $64$-bit, the maximum length of data that could be processed in one (assembly) instruction execution varies between $4-8$ bytes.
Hence, the elements within $D$ may or may not fit a single instruction within the malware trace.
For instance, an IPv4 address could be matched in a single instruction unlike a long string value that would require multiple consecutive instructions.

Hence, in this step, we introduce two strategies to identify \acp{POI}: \emph{Standalone \ac{POI}} and \emph{Contiguous \ac{POI}}.
As the name implies, the first strategy handles \ac{POI} identification for values that fit a standalone (single) instruction.
The second strategy handles \acp{POI} identification for values that cannot be processed within a single instruction.

\subsubsection{Standalone \ac{POI}}
This strategy attempts to identify \acp{POI} related to data $d \in D$ which can be processed by a single instruction.
There are three possible ways such an instruction could interact with $d$:
\begin{itemize}
    \item writing to memory, e.g., a \lstinline{push eax} instruction
    \item reading from memory, e.g., a \lstinline{mov eax, [esp]} instruction
    \item reading from or writing to a register
\end{itemize}

Hence, this strategy loops over each line of the trace file and compares the registers, memory reads, and memory writes with every $d \in D$.
Whenever data access of an instruction in the trace matches some $d\in D$, the corresponding instruction's address is marked as a \ac{POI} candidate.

\subsubsection{Contiguous \ac{POI}}
\label{sec:poi_mem_pattern_poi}
This strategy is responsible to handle scenarios when data $d \in D$ requires more than one consecutive instruction for handling it.
In comparison to the previous scenario, multi-instruction scenarios are more complicated because the order how data $d$ is accessed could be sequential, reversed, or even random.
Hence, this strategy needs to iteratively search for an expected pattern, i.e., $d$, after each memory write-operation.

\begin{figure}
    \centering
    \includegraphics[scale=.8]{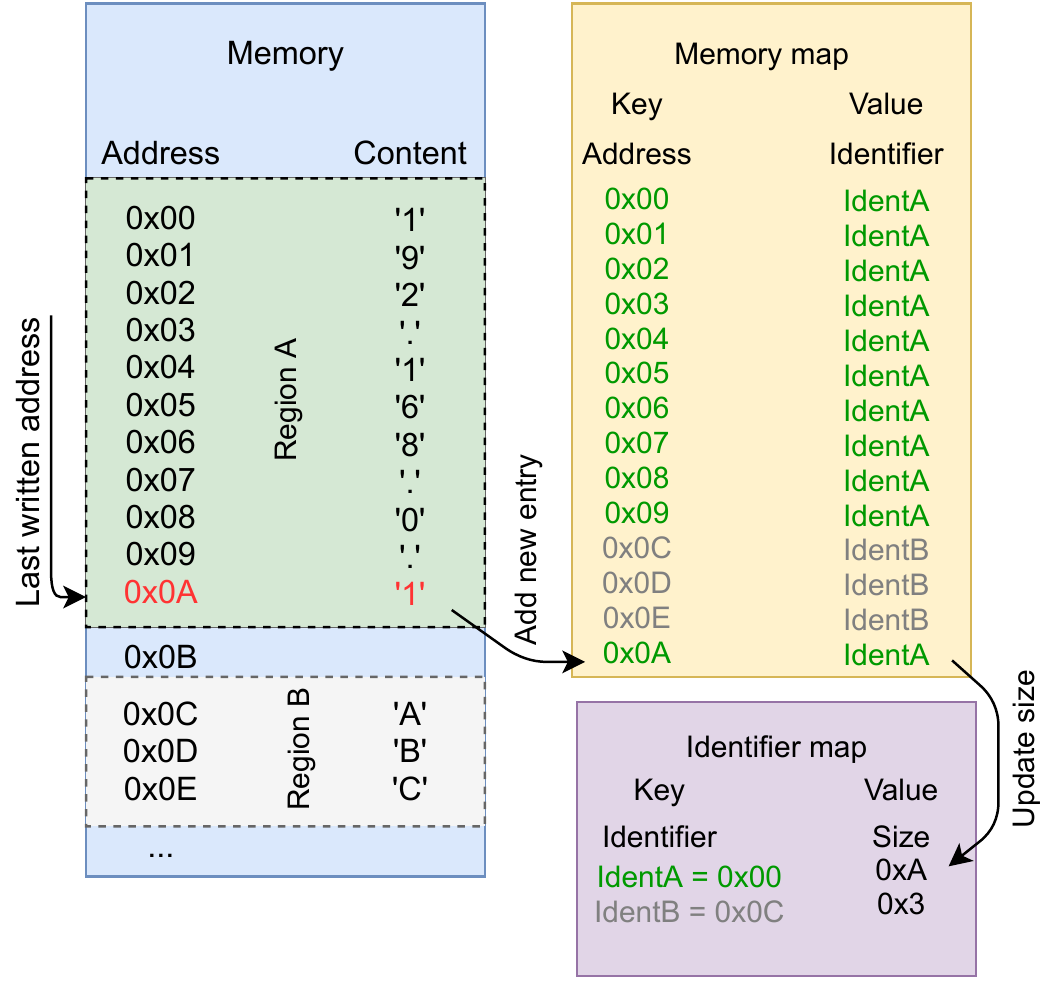}
    \caption{Tracking of memory block utilization using lookup tables}
    \label{fig:memory-map} 
\end{figure}

In order to make the search for the pattern more efficient, the memory should not be searched exhaustively. 
Only the memory written by the malware should be searched.
However, the memory allocation information by the \ac{OS} only reports the allocation to an application or process. 
It does not inform on the exact utilization of the block.
Hence, we needed the ability to track the utilization of the allocated memory by the malware process, i.e., actual memory writes.
For this purpose, we leveraged lookup tables as illustrated in Figure \ref{fig:memory-map} to facilitate our tracking purposes.

\paragraph{Lookup Tables}
To track the memory write operations for a process, we need to introduce two custom lookup tables:\textit{Memory Map} and \textit{Identifier Map} .
For each memory block assigned to the malware process, we need to keep track of all write operations, i.e., content written, during \emph{runtime}.
Using the Memory Map, a \textit{memory region} should be dynamically assigned and tracked such that each address within a contiguous block of written memory is mapped to its corresponding region.
To effectively track memory regions, all memory write-operations from the trace file must be considered with an assumption that the file includes all execution details of the malware, right from the beginning.
Finally, the Identifier Map should be used to keep track the size of each memory region being tracked.
Hence, for each change recorded in the Memory Map, the corresponding new size of the affected memory region should also be updated accordingly in the Identifier Map. 

\paragraph{Memory Pattern Search}
As mentioned earlier, for each memory write-operation recorded within the malware trace file, we need to search if any memory region in the memory matches the patterns of $d \in D$.
Leveraging the lookup tables introduced previously, for each accessed memory address, a memory pattern search is executed for each $d \in D$ based on \autoref{alg:searching-pattern}.
First, the information of the memory region for the currently accessed memory address is retrieved from the lookup tables (Line $2-5$).
Then, the algorithm checks if the considered pattern $d$'s size is smaller or equal the size of the currently considered memory region (Line $6$).
If the size of the pattern fits the considered region, the contents within the memory region is compared to the pattern of $d$ (Line $7-8$). 
If a match was successfully found, the algorithm returns the address of the match, i.e., as a \ac{POI} candidate.

\begin{algorithm}
    \SetKwData{true}{true}
    \SetKwData{return}{return}
    \SetKwData{in}{in}
    \SetKwData{findMemoryRegion}{findMemoryRegion}
    \SetKwData{address}{lastAccessedAddress}
    \SetKwData{size}{size}
    \SetKwData{pattern}{pattern}
    \SetKwData{identifierMap}{identifierMap}
    \SetKwData{memoryMap}{memoryMap}
    \SetKwFunction{min}{min}
    \SetKwFunction{max}{max}
    \SetKwFunction{find}{find}
    \SetKwFunction{readMemory}{readMemory}

    \KwIn{\address, \pattern}
    \KwResult{The start address of the searched pattern \pattern if it is found. Otherwise, -1 is returned.}

    \BlankLine
    // A region identifier is the start address of a memory region
    
    $\textit{regionIdentifier} \gets \memoryMap[\address]$
    
    $\textit{regionSize} \gets \identifierMap[\textit{regionIdentifier}]$
    
    \textit{startAddr} $\gets$ \textit{\max{regionIdentifier, \address - \pattern.size}}
            
    \textit{endAddr} $\gets$ \textit{\min{regionIdentifier + regionSize, \address + \pattern.size}}
            
    \If{$\pattern\text{.size} \leq \textit{(endAddr-startAddr)}$}
        {
            $\textit{buffer} \gets \readMemory{\textit{startAddr}, \textit{endAddr}}$
            
            $\textit{pos} \gets \textit{buffer}.\find{\pattern}$
            
            \return \textit{pos};
        }
    \return -1;
    
    \caption{Searching a Pattern in Memory}
    \label{alg:searching-pattern}
\end{algorithm}

\label{sec:pinpuppet_sys_overview}
\begin{figure*}
    \centering
    \includegraphics[scale=0.9]{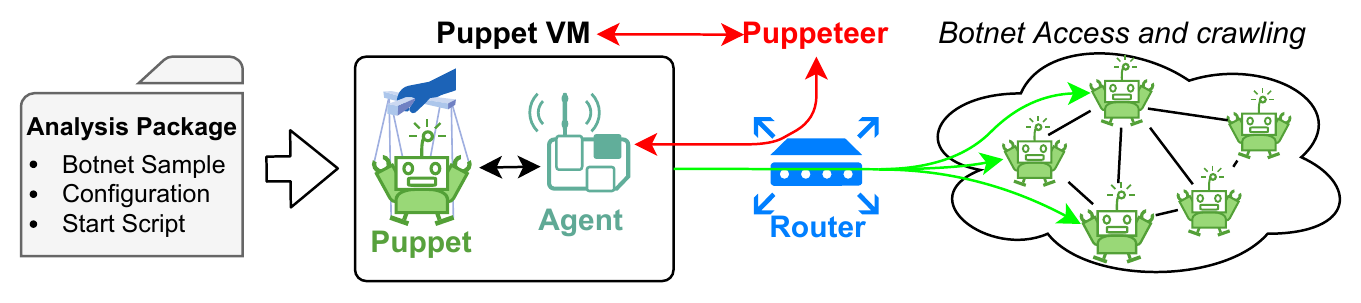}
    \caption{System overview for PinPuppet.}
    \label{fig:sys_overview}
\end{figure*}

\subsection{Filtering \acp{POI}}
\label{sec:poi_ensuring_the_quality_of_pois}
The \ac{POI} candidates identified in the previous section are merely pointing to addresses that were used to access data $d \in D$ during execution.
However, due to the nature of low quality values of $d$, i.e., non-unique, or simply due to chance, not all \acp{POI} are suitable as beacons, i.e., may introduce false positives.
This is also highly dependent on the exact use case for the \acp{POI} as beacons. 
If they are used only for guiding manual reverse engineering, low quality \acp{POI} can be easily ignored by a skilled analyst. 
For instance, a \ac{POI} candidate might be flagged due to the address referring to a general purpose memory copying function, e.g., \texttt{memcpy}.
This may or may not be of interest to the analyst, i.e., information about the flows of data in $D$ within the malware binary. 
However, for tools that depends on the \acp{POI} for further analysis or those that rely on \ac{POI} instructions that are known to always access some data from $D$, a low-quality \acp{POI} could be detrimental (see \autoref{sec:pinpuppet}).

As the quality of a \ac{POI} is not an absolute measure and the fact that determining the quality is only possible with manual reverse engineering, we introduce \textbf{confidence score} as a metric to estimate the quality of a \ac{POI}. 
This metric assesses the frequency of an address associated to a particular \ac{POI} being used to access \textit{only} elements of $d \in D$.
By assigning a score between $[0.0;1.0]$ for each \ac{POI}, the quality of the address can be estimated.
A score of $1.0$ would indicate that the \ac{POI} is used exclusively by the malware for accessing only data in $D$.
Meanwhile, scores of $<1.0$ indicate that other contents (not in $D$) are also accessed using the address of \ac{POI}.

In the following, we elaborate the details of the metric which  needs to be adapted accordingly to the search strategies outlined in \autoref{sec:poi_identify}, i.e., Standalone and Contiguous \acp{POI}.

\subsubsection{Standalone \acp{POI}}
In this section, the adaptation of the confidence score metric for Standalone \acp{POI} is described based on \autoref{alg:confidence_score_calc}.
The confidence score of each Standalone \ac{POI} (Line $9$) can be expressed as the ratio of instructions within the malware binary that processes data from $D$ (Line $8$) compared to all data processed by those instructions (Line $4$). 

The confidence score of these kind of \ac{POI} is most reliable when the malware is iterating a command with multiple different values, e.g., conducting DDoS attack with multiple spoofed addresses.
Once the confidence score for the \acp{POI} are calculated, a certain threshold can be used to provide a cutoff in selecting high-quality \acp{POI} as beacons. 
Discussion on how to select the threshold value is provided in \autoref{sec:evaluation_confidence_score_estimator}.

\begin{algorithm}
    \SetKwData{poi}{poi}
    \SetKwData{trace}{trace}
    \SetKwFunction{append}{append}
    \SetKwFunction{len}{len}
    \SetKwData{return}{return}

    \KwIn{\poi, \trace, $D$}
    \KwResult{The confidence score $x\in[0.0;1.0]$}

    \BlankLine
    \textit{data\_sequence} $\gets$ $[]$
    
    \For{trace\_entry in \trace}{
        \If{trace\_entry.address = \poi.address}{
            \textit{data\_sequence}.\append{\textit{trace\_entry.processed\_data}}
        }
    }
    
    \textit{correct\_counter} $\gets 0$
    
    \For{$d$ in data\_sequence}{
        \If{$d\in D$}{
            \textit{correct\_counter} $\gets \textit{correct\_counter}+1$
        }
    }
    
    \return $\dfrac{\textit{correct\_counter}}{\len{\textit{data\_sequence}}}$\;
    
    \caption{Confidence score of Standalone \acp{POI}.}
    \label{alg:confidence_score_calc}
\end{algorithm}

\subsubsection{Contiguous \acp{POI}}
The adaptation of the confidence score metric for Contiguous \acp{POI} is not as straightforward as it was for Standalone \acp{POI}.
For Contiguous \acp{POI}, a mapping from memory address to the \textit{last} instruction address that wrote to the memory address is required to determine the confidence score.
This can be achieved through the lookup tables introduced in \autoref{sec:poi_identify}.

In more detail, let $P$ be a list of sequences of instruction addresses that wrote the bytes of the found patterns $\textit{$ins_1, ..., ins_n$}$ where $ins_i$ wrote the i\textsuperscript{th} byte of a found pattern. 
Then, let $C(x, S)$ be the number of elements of a sequence $S = s_1, ...,s_n$ that are equal to $x$. 
In addition, let $\#poi_i$ be the number of bytes a \ac{POI} writes in the runtime of the process. 
With that, the confidence score for a \ac{POI} $poi_i$ is calculated using 
$$\sum_{p}^{P} \dfrac{C(poi_i, p)}{\#poi_i}$$

Simply put, for a \ac{POI}, we calculate the number of pattern bytes written, divided by the total number of bytes it wrote. 
It is important to note that for pattern bytes that are written, we count only the bytes that occurred in a complete pattern, i.e., full. 
This is required to prevent an instruction that only writes parts of the pattern from not receiving a high score.
The confidence score for a read-memory pattern \acp{POI} is also calculated analogously.

Unlike Standalone \acp{POI}, the presence of overlapping data $d \in D$ is not much of a problem. 
This is because data that cannot be processed within one instruction is usually much longer. 
So the probability that an ASCII string "192.168.0.1" is not an IP but some other data is much lower than if it is represented as an integer, e.g., 3232235521. 
However, this is only valid for data that are lengthy and not shorter ones like the representation of a port (decimal) value $8$ in ASCII.
Once again, after obtaining the relevant confidence scores for all \acp{POI}, a threshold value can be selected to filter out only high-quality \acp{POI} as beacons. 
Please refer to \autoref{sec:evaluation_confidence_score_estimator} for discussion on choosing the suitable threshold value.

After filtering out the high-quality \acp{POI}, malware analysts can integrate the beacons into their reverse engineering workflows.
To assist with that, we have additionally implemented two plugins for \textit{IDA Pro} and \textit{Ghidra} that can highlight the instructions that accesses the \acp{POI} for the given malware sample.
\autoref{sec:appendix_plugins} provides further details about the plugins.

Even though malware analysts benefits from using our proposed \acp{POI} as beacons for manual reverse engineering, this is not the only use case. 
Our \acp{POI} can also be used by more advanced analysis tools to simplify the analysis of a malware sample. 
In the next section, we introduce PinPuppet as a tool that leverages our \acp{POI} to automatically monitor P2P botnets.

%% file: 040_pinpuppet.tex
\section{PinPuppet Monitoring}
\label{sec:pinpuppet}
Based on our approach from \autoref{sec:poi} we developed a proof-of-concept prototype called PinPuppet. 
PinPuppet extends the \ac{POI} mechanism with additional components as described in \autoref{sec:pinpuppet_sys_overview} to enable features to automatically instrument a malware to conduct botnet monitoring.
The overall goal of PinPuppet is to crawl a P2P botnet without requiring detailed knowledge about the sample or the botnet itself, e.g., the MM-protocol.

Our prototype can be used to evaluate whether \acp{POI} can be used for advanced malware analysis and to verify that the methodology for identifying and filtering \acp{POI} works. 
We decided to go on with this evaluation strategy in comparison to a more subjective evaluation of how the beacons are useful for malware analysts.

The scope of PinPuppet is limited to botnets which do not rely on destination-dependent messages, e.g., encrypted with a recipient's public key. 
Furthermore, PinPuppet assumes that the malware sample is instrumentable by Intel Pin (in the following only Pin), the choice of our \ac{DBI} framework. 
Consequently, all anti-Pin~\cite{kirschPwINPwningIntel2018a,deliaSoKUsingDynamic2019,gorgovanEscapingDynamoRIOPin2021} mechanisms of the sample are assumed to be circumvented.

The information $D$ in the form of bootstrap list $\BS$ of a sample is assumed to be known and access to a \ac{VM} with snapshot functionality is available. 
The former can be done automatically, e.g. by running the sample on a sandbox and logging all peers contacted by it. 
In addition, it is assumed that all anti-\ac{VM} countermeasures of the malware (if applicable) are also circumvented.

\subsection{System Overview}

PinPuppet is not a monolithic system. It consists of multiple components, which each fulfills their own purpose. 
Considering that PinPuppet implements crawling functionality, we need to repeatedly evoke the \textit{getL} function (cf. \autoref{sec:bg_p2p_botnets}). 
As this is typically one of the first messages sent after the initial connection, we could achieve this by restarting the sample, however, two problems arise: Firstly, the duration from restart to performing the connection could be relatively long (e.g., approximately 5 minutes for Sality) and, secondly, \acp{POI} would not stay valid throughout consecutive restarts.

\acp{POI} reference the addresses of instructions and thus, they only stay valid as long as the sample's memory layout is not reorganized or the content is not changed. This requires that neither \ac{ASLR} nor any malware anti-reverse engineering mechanism that modifies the corresponding instruction address. This can be achieved by relying on virtualization and live snapshots to keep a state where the \acp{POI} reference to the corresponding instructions in memory is retained.

Thus, we utilize a \ac{VM} with a ``live'' snapshot where the sample is already running. This prevents ASLR or even the malware itself from randomizing the memory layout.
The interaction between the various components within PinPuppet is illustrated in \autoref{fig:sys_overview}.
Each component is detailed individually in the following.

An \textbf{Analysis Package} bundles all information that is needed for crawling a botnet. It contains the malware sample, a configuration file, and a script that describes how to start the sample and how to attach Pin to the running malware. The configuration contains the bootstrap list, and the values $\TT$ and $\TC$  which will be explained later on.

The \textbf{Puppet VM}, as the name suggests, is a virtual machine (Windows 7 32-bit). The two main components that are running on this VM are the \textbf{Puppet} and the \textbf{Agent}.  The Agent is our foothold in the Puppet VM. It allows uploading analysis packages, running them, and downloading the results.
Meanwhile, the Puppet is the botnet sample with Pin attached to it. Pin is then used to trace instructions based on filtering criteria -- we will elaborate those later on in \autoref{sec:pin_puppet_step_trace_data_collection}. Additionally, the Puppet wraps calls to the Windows socket API (winsock2.h\footnote{\url{https://docs.microsoft.com/en-us/windows/win32/api/winsock2/}}).
We call this component the Socket Wrapper. Whenever a sending API function is called, the IP and port parameters are first sent over the network using the Agent. The Socket Wrapper then waits for a response (which contains no data) before executing the original API call. For receiving API calls, the order is reversed, i.e., the API call is executed before the results are sent using the Agent. The Socket Wrapper again waits for a reply before continuing. By not replying to such a message, we can thus pause a VM immediately before performing a sending network operation or after performing a receiving network operation.

The \textbf{Router} is used for routing the traffic between the Puppet VM, the Puppeteer, and the botnet itself. Before crawling a peer, the Router is reconfigured by the Puppeteer. Afterwards, the configuration is rolled back to allow crawling the next peer.

The \textbf{Puppeteer} is the brain of PinPuppet. It sets up the Router, controls the Puppet VM (e.g., creating snapshots, starting, and stopping it). It communicates with the Agent via the network and sends commands to it. Data collected by the Puppet, is downloaded by the Puppeteer using the Agent for further processing.

\subsection{Methodology}
The automated crawling performed by PinPuppet is a five step process. The first step sets up a snapshot where the sample is running. As explained in the previous section, this is required to make sure that the memory layout does not change and thus our \acp{POI} stay valid throughout the whole process. Steps two, three and four correspond to the equally named steps from the \acp{POI} beacon methodology (cf. \autoref{fig:poi_discovery}). The process explained in the final step then allows one to a) contact a user specified peer and b) extract peers that were shared by the contacted peer in \textit{getL} replies. This can be used to crawl the botnet. In the following, we will explain the individual steps in detail.

\subsubsection{Snapshot Creation}
\label{sec:pin_puppet_step_snapshot_creation}
In this step, PinPuppet firstly uses the Socket Wrapper to pause the sample by pausing the \ac{VM} directly before it initiates the connection to a peer and, secondly, creates a snapshot of the VM where the sample is in the paused state. We only pause the sample when the address it is trying to contact is in $\BS$. This ensures, that it is contacting a real peer and, for example, not a multicast or local address. We call the peer it is about to contact the original peer $p_o$. The sample is kept disconnected from the botnet in this step and thus never establishes a connection. For UDP, we pause the malware before calling a \lstinline{sendto}-family API call. For TCP, we pause the malware before calling a \lstinline{connect}-family API call. The snapshot that results from this process is called ``running snapshot.''

\subsubsection{Trace Data Collection}
\label{sec:pin_puppet_step_trace_data_collection}
As a requirement for extracting \acp{POI} is the existence of an instruction trace (cf. \autoref{fig:poi_discovery}), we record such a trace in this step. For this recording step, the Puppet VM has access to the Botnet, i.e., it will exchange MM-messages with bots in the botnet. It is then restored to the running snapshot and where the sample is resumed. The sample is then left running for a fixed amount of time $\TT$ (configured using the Analysis Package) to allow for the bot to perform MM-communications with the botnet. We have used a duration longer than the length of one MM-cycle. The length of one MM-cycle can either be taken from existing literature or by analyzing the traffic generated by the sample (e.g., using \textit{Cuckoo Sandbox}). In general, running the sample for longer is always an option which only improves the quality of the results. However, the processing time required in the next two steps will increase with the amount of data collected in this step.

While the botnet is running, it generates trace data and Socket Wrapper logs. After time $\TT$ has elapsed, the Puppeteer downloads the generated data from the VM and, lastly, shuts it down. The Socket Wrapper logs are of particular interest to us, because they give us the information of which peers were contacted by the Puppet (outgoing) and which peers it was contacted by (incoming). The set of all peers identified using the Socket Wrapper logs is called $\SW$.

Tracing all instructions with Pin can introduce a significant runtime overhead for the application being traced and creates enormous amounts of data. Since the tracing of programs becomes so slow that it is no longer practical, we have added various filtering options. They allow the user to more closely specify the instructions which should be traced. The implemented filtering options are:
\begin{itemize}
    \item whether instructions belonging to named regions should be traced and, if so, for which regions.
    \item whether instructions not belonging to named regions should be traced (e.g., a memory region where the sample stores unpacked code).
\end{itemize}
In addition, one can specify the maximum number of times an instruction is traced -- the maximum trace count. This prevents instructions which are executed very often from unnecessarily bloating the generated trace data with useless information. We have observed this behavior with Sality, which, without the maximum trace count, creates a 30GB trace containing mostly data from a small set of instructions. These filtering options are also specified as part of the analysis package.

\subsubsection{POI Candidate Identification}
\label{sec:pin_puppet_step_poi_candidate_identification}
In this step, the Puppeteer performs the \ac{POI} candidate identification (cf. \autoref{sec:poi_identify} and \autoref{fig:poi_discovery}) for identifying IP- and port-\acp{POI}, i.e., \acp{POI} corresponding to instructions frequently processing IPs or ports. Port-\acp{POI} are especially important for botnets using varying listening ports, however, for botnets with a fixed listening port, Port-\acp{POI} can be skipped as they would provide no additional information. 

The identification process uses the data set $D_i$ for the IP-\acp{POI} and, if applicable, the data set $D_p$ for port-\acp{POI}. Both data sets are constructed based on the IPs and ports in $\BS\cup\SW$. For both the IPs and the ports, the following representations are added to $D_i$ and $D_p$, respectively:
\begin{itemize}
    \item 4-byte binary \ac{LSB}-first and 4-byte binary \ac{MSB}-first (used for Simple-\ac{POI} identification).
    \item ASCII representation, used for identifying memory pattern \acp{POI}. 
\end{itemize}
The intuition behind using $\BS\cup\SW$ as the basis for our data sets is that it approximates all peers that the puppet can know. The true set of known peers is $\BS\cup\SW\cup\PS$ where $\PS$ are all peers that were shared with the Puppet in the trace data collection step. Of course, we do not know which peers were shared with the Puppet, however, as bots who request new shared peers oftentimes contact these shared peers, there is an overlap between $\SW$ and $\PS$. Thus the difference between $\BS\cup\SW\cup\PS$ and $\BS\cup\SW$ is relatively small.

\subsubsection{POI Filtering}
\label{sec:pin_puppet_step_poi_filtering}

\begin{algorithm}
    \SetKwData{IPPois}{IPPois}
    \SetKwData{PortPois}{PortPois}
    \SetKwData{ipPoi}{ipPoi}
    \SetKwData{extractedIps}{extractedIps}
    \SetKwData{portPoi}{portPoi}
    \SetKwData{ip}{ip}
    \SetKwData{port}{port}
    \SetKwData{extractedPorts}{extractedPorts}
    \SetKwData{poiMapping}{poiMapping}
    \SetKwData{extractedIPs}{extractedIPs}
    \SetKwData{extractedPorts}{extractedPorts}
    \SetKwData{true}{true}
    \SetKwData{false}{false}
    \SetKwData{matches}{matches}
    \SetKwFunction{append}{append}
    \SetKwFunction{zip}{zip}

    \KwIn{\IPPois, \PortPois, $\BS\cup\SW$}
    \KwData{PoiMapping<IPPoi, List<PortPoi>{}>: \poiMapping}
    \KwResult{The mapping between IP- and port-\acp{POI}.}

    \BlankLine
    
    // \textit{extractedIPs} and \textit{extractedPorts} correspond
    
    // to the data processed at the respective \ac{POI}.

    \For{\textit{ipPoi}, \textit{extractedIPs} in \IPPois}{
        \For{\textit{portPoi}, \textit{extractedPorts} in \PortPois}{
            \textit{matches}$\gets$\true
            
            \For{ip, port in \zip{\extractedIps, \extractedPorts}}{
                \If{$(\textit{ip},\textit{port})\not\in\BS\cup\SW$}{
                    \textit{matches}$\gets$\false
                }
            }
            \If{\textit{matches}}{
                \poiMapping[\textit{ipPoi}].\append{\textit{portPoi}}
            }
        }
    }
    
    \caption{Matching IP- and port-\acp{POI}.}
    \label{alg:matching_pois}
\end{algorithm}
In this step, the Puppeteer performs \ac{POI} filtering (cf. \autoref{sec:poi_identify} and \autoref{fig:poi_discovery}) for the IP-\acp{POI} identified in the previous step. The puppeteer calculates the confidence score of each IP-\ac{POI}. Using the confidence score threshold, the IP-\acp{POI} are filtered. In this case, confidence score threshold is required to compensate for the aforementioned difference between $\BS\cup\SW\cup\PS$ and $\BS\cup\SW$.

For port-\acp{POI} (if applicable), we perform a different process: as we want to extract IP and port tuples from traces in the next step, we need to find the mapping between IP- and port-\acp{POI}. To obtain the mapping, we first, for every \ac{POI}, extract all IPs and ports using the \acp{POI} from the trace data generated in the second step. We then use \autoref{alg:matching_pois} to create the mapping. All port-\acp{POI} which are not in the resulting mapping are discarded (regardless of confidence score).

\subsubsection{Crawling Primitive}
\label{sec:pin_puppet_step_crawling_primitive}
In this step, the crawling is performed. In the following we present the primitive that is used to a) contact a specified ``crawl peer'' $p_c$, and b) extract peers shared by $p_c$.

First, the router is configured to redirect outgoing traffic destined to $p_o$ to $p_c$. For any replies sent by $p_c$ the Router is configured to change the source to $p_o$. Thus, the Puppet assumes it is talking to $p_o$ even though it is communicating with $p_c$.

To prevent unsolicited requests to the Puppet, e.g., because it previously contacted $p_x$ which now tries to probe the puppet, all traffic that is not between the Puppet and $p_c$ is blocked. This is to make sure, that the extracted peers are only based on the short communication with $p_c$.

The sample is now left running for a short time period $\TC$ (configured using the Analysis Package) after which the trace data is downloaded. Using the IP- and port-\acp{POI}, the Puppeteer extracts IPs and ports from the trace data. For every IP- and port-\ac{POI} which are mapped together in the \ac{POI} mapping, the extracted IPs and ports are combined to form full addresses.

Finally, the Puppet VM is stopped and all settings made on the router are reverted. The resulting address list is returned as the result of the primitive. 

The primitive introduce in this step can then be used to implement a fully fledged crawler, for example, using \ac{BFS} or \ac{DFS}~\cite{rossowSoKP2PWNEDModeling2013}, or more advance algorithms such as the \ac{LICA}~\cite{karuppayahAdvancedMonitoringResilient2014}. PinPuppet implements a \ac{BFS} crawler using this primitive.

\subsection{Monitoring Live Botnets}
When running PinPuppet with a live botnet on the Internet, additional measures need to be taken to ensure that the Puppet employed by PinPuppet does not contribute to the malicious activity of a botnet. This is especially relevant in the trace data collection step (cf. \autoref{sec:pin_puppet_step_trace_data_collection}) where the Puppet is left running for a longer duration. 
In these cases we suggest measures like blocking relevant ports (e.g., SMTP), significantly limiting the available bandwidth to reduce the possible participation in DDoS attacks, and reducing the time $\TT$. When using the crawling primitive (cf. \autoref{sec:pin_puppet_step_crawling_primitive}), the problem is still relevant, however, the problem is reduced as the sample is only running for shorter periods of time before being reset. In addition, the router configuration in the primitive prevents any access to peers except the one being crawled. Overall, before using PinPuppet to monitor a live botnet on the Internet, a botnet should be studied thoroughly to ensure proper containment mechanisms are set up.

%% file: 060_evaluation.tex
\section{Evaluation}
\label{sec:evaluation}

In the evaluation, we want to answer the question of whether PinPuppet is able to leverage \acp{POI} to crawl P2P botnets. For this, we will analyze multiple aspects of our PinPuppet implementation using multiple tailored analysis packages for several botnets. For each of these botnets we have a local botnet set up to provide real-world MM-communications.

For the foundation, we want to analyze whether the execution overhead introduced by PinPuppet affects the sample's speed. This is important, as too high of an overhead can lead to the sample crashing or altering it behavior, e.g., due to transmission timeouts.

The next part of our evaluation targets the number, category, and confidence score of \acp{POI} identified by PinPuppet. For our crawling primitive to work, we need to identify \acp{POI} that have a high quality, i.e., confidence score (cf. \autoref{sec:poi_ensuring_the_quality_of_pois}). Without high quality \acp{POI}, we would not be able to correctly extract peers in the crawling primitive. In addition, the more \acp{POI} PinPuppet identifies, the more information a reverse engineer has for future manual reverse engineering. 

As we use the confidence score to predict the quality of an IP-\acp{POI}, it is important that the confidence score is a good estimator for this quality. The quality of a \ac{POI} directly affects the quality of the IPs extracted by it. On the one hand, if the confidence score were to overestimate the quality, we would extract spurious IPs using the \acp{POI}. On the other hand, if the confidence score were to underestimate the quality, we would disregard good \acp{POI} and potentially miss IPs which we could have extracted using these \acp{POI}. Thus, we analyze how well of an estimator the confidence score is.

The last part of our evaluation focuses on the amount and types of extracted peers, i.e., on how effective PinPuppet is in crawling a botnet. 
This is a crucial aspect, as the ability to crawl a botnet, would also indirectly validate the methodology for \acp{POI} identification and prove their usefulness as beacons for reverse engineering.

\subsection{Evaluation Set}
\label{sec:evaluation_set}
For the evaluation of PinPuppet and \acp{POI} we decided not to rely on live botnets, as we needed a reproducible environment. Instead we bootstrapped botnets -- using real-world malware samples -- locally, i.e., isolated from the Internet. For the identification of \acp{POI}, this has no negative impact, because the bot will not realize that it is only talking to a local botnet. Thus it behaves the same and uses the same instructions as if it were talking to a real botnet. 

Our used approach is similar to the one used by \textit{Calvet et al.} for creating an in-the-lab Waledac botnet~\cite{Calvet2010}. However, their approach includes many aspects of a botnet, such as the \ac{C2}, and thus requires significant reverse engineering efforts. Our approach, in contrast, is more general and focuses on the MM-mechanism and -protocol. It thus needs less reverse engineering, requiring only the bootstrap list as prior knowledge of a sample. 
Our process for bootstrapping such a \textit{local botnet} with $n+m$ peers (denoted by the set $\LP$) is as follows:
\begin{enumerate}
    \item Create $n$-VMs with IPs from the bootstrap list and run the sample on the VMs. We call these peers the ``local bootstrap peers.''
    \item Wait for the local bootstrap peers to establish a local botnet. In case of $n=1$, one needs to wait for the peer to be fully initialized.
    \item Create $m$-VMs with arbitrary public IP addresses. These peers will contact the local bootstrap peers and join the botnet overlay network.
\end{enumerate}
After this process, we use live snapshots to store the state of the botnet and, later on, restore the state when running PinPuppet.

For the evaluation, we use four botnets: ZeroAccess~\cite{nevilleZeroAccessIndepth2013,karuppayah2018advanced},
Sality~\cite{karuppayah2018advanced},
Nugache \cite{stoverAnalysisStormNugache2007}
, and Kelihos \cite{kelihos}.

For Nugache, we have $n=8$ local bootstrap peers and $m=40$ other peers. For the other botnets we have $n=1$ local bootstrap peer and $m=40$ other peers. The hashes for the samples used are given in the \autoref{sec:appendix_hashes}. ZeroAccess and Sality utilize UDP. Nugache and Kelihos utilize TCP.

The Nugache local botnet requires eight local bootstrap peers, as every peer has at most 15 active connections~\cite{stoverAnalysisStormNugache2007}. Thus, only having one local bootstrap peer is not enough. Not all peers who join the overlay receive new local peers.

Sality does not use a fixed port and, instead, uses ports which depend on the computer name~\cite{karuppayah2018advanced}. To make our Sality local botnet more realistic, all peers in the local botnet have unique computer names. In addition, we have added NAT rules to set the correct port for MM-traffic being sent to the local bootstrap peer.

As additional data for PinPuppet we have identified the bootstrap list for each botnet (cf. \autoref{sec:pinpuppet}). For ZeroAccess our identified bootstrap list contains 256 peers, for Sality 740, for Nugache 24, and for Kelihos 316. The exact configuration, i.e., the analysis package, used for each botnet and PinPuppet is available in our Git repository\footnote{The git repository will be made available for the camera-ready version of this paper.}.

\subsection{Experiments and Results}
\label{sec:results}
In the following, we will present our evaluation results and an analysis. In the analysis, we will cover the aspects mentioned in \autoref{sec:evaluation}, namely, that one can use PinPuppet and \acp{POI} to crawl a P2P botnet. The results were gathered by running PinPuppet for the evaluation set (cf. \autoref{sec:evaluation_set}) while having access to the corresponding local botnet.
\subsubsection{How does the execution overhead introduced by PinPuppet affects the sample's speed?}
\label{sec:eval_pinpuppet_speed}
For analyzing the overhead, we run each sample five times with and five times without Pin attached to it. As metric we use the execution time with and without PinPuppet for the points $T_0$, $T_1$, and $T_2$. At $t=0$ the sample is started. $T_0$, $T_1$, and $T_2$ are the times where the sample sends out the first, 20\textsuperscript{th}, and 40\textsuperscript{th} message, respectively.

As expected, the differences between $T_1$ and $T_0$ or $T_2$ and $ T_1$ with and without PinPuppet are minimal. The largest average difference measured for these metric is a 6.5\% increase for the duration from $T_0$ to $T_1$ for Sality. Botnet malware are not necessarily designed as time-critical applications. Especially after the initialization when the botnet sample is contacting peers from its neighbor list, i.e., after $T_0$, the overhead is minimal as this phase involves a lot of waiting on the samples side. Additionally, our filtering mechanisms (cf. \autoref{sec:pin_puppet_step_trace_data_collection}) limit the number of instrumented instructions and thus reduce the overall overhead.

In the duration until contacting the first peer, i.e., until $T_0$, we have observed that PinPuppet incurs a significant runtime overhead. The most extreme example is Sality that, coupled with PinPuppet, reaches $T_0$ at $T_0=689.36s$ and at $T_0=335.58s$ without PinPuppet. We speculate that this is due to the complex unpacking process happening in this first interval, and, depending on the trace filter settings, the instructions for this unpacking are being traced, introducing the overhead. The full table with our results can be seen in \autoref{sec:appendix_overhead}.

\subsubsection{How many and which \acp{POI} are identified by PinPuppet and what is their respective confidence score?}
\begin{figure*}%
    \centering
    \includegraphics[scale=0.16]{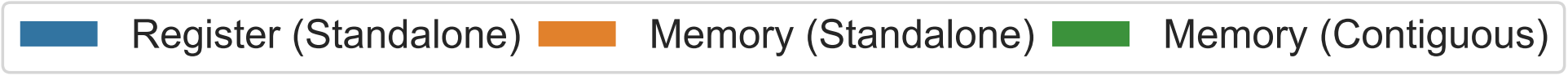}
    \subfigure[ZeroAccess]{%
        \includegraphics[scale=0.44]{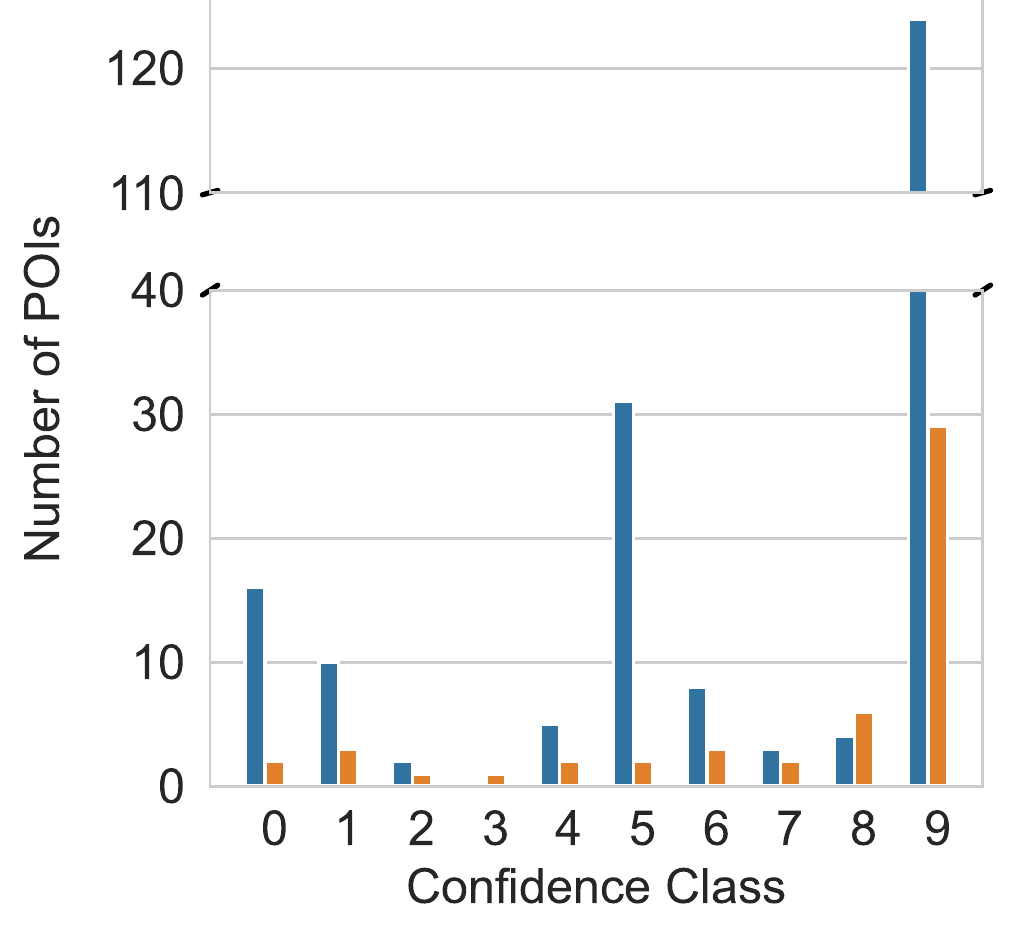}%
    }%
    \subfigure[Sality]{%
        \includegraphics[scale=0.44]{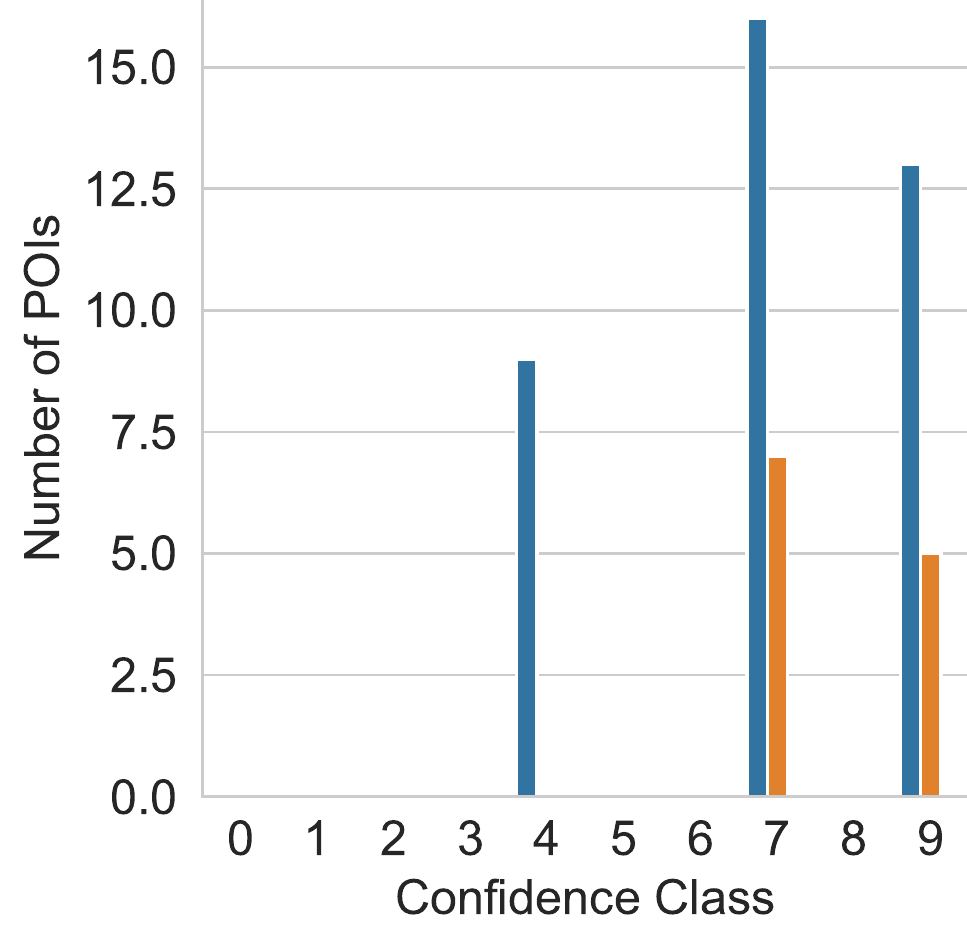}%
    }%
    \subfigure[Nugache]{%
        \includegraphics[scale=0.44]{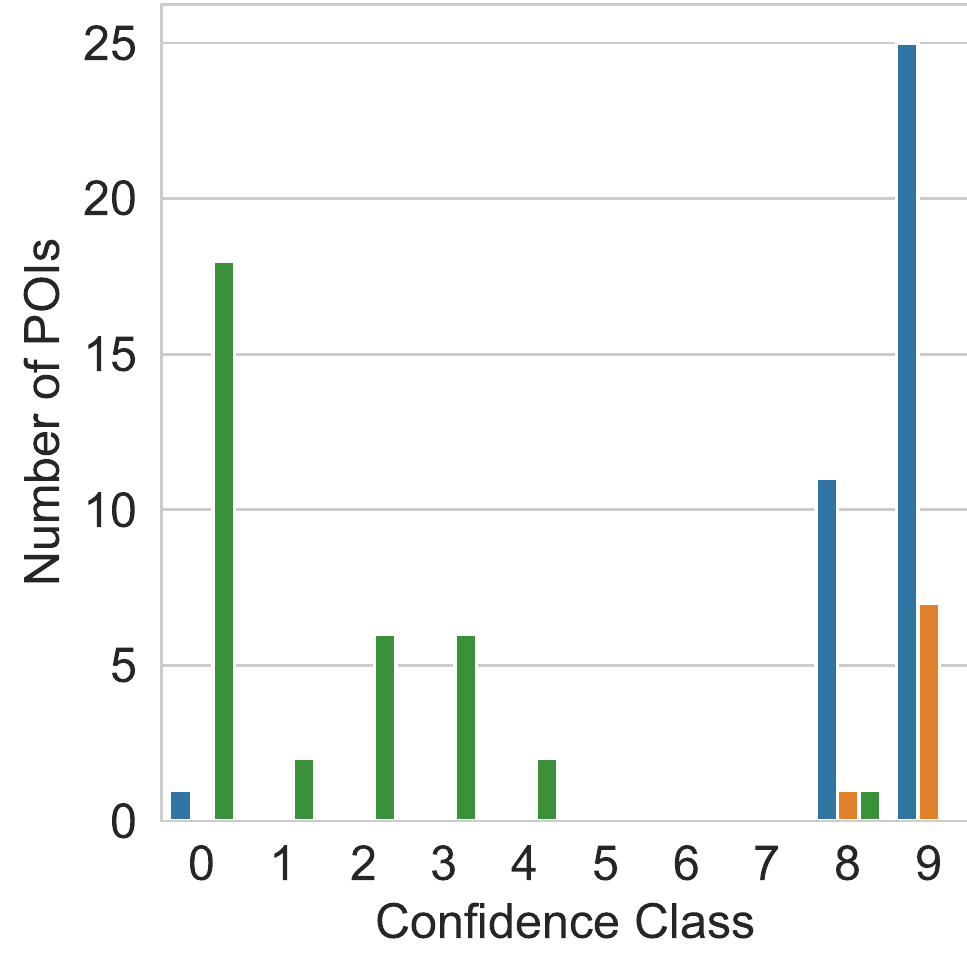}%
    }%
    \subfigure[Kelihos]{%
        \includegraphics[scale=0.44]{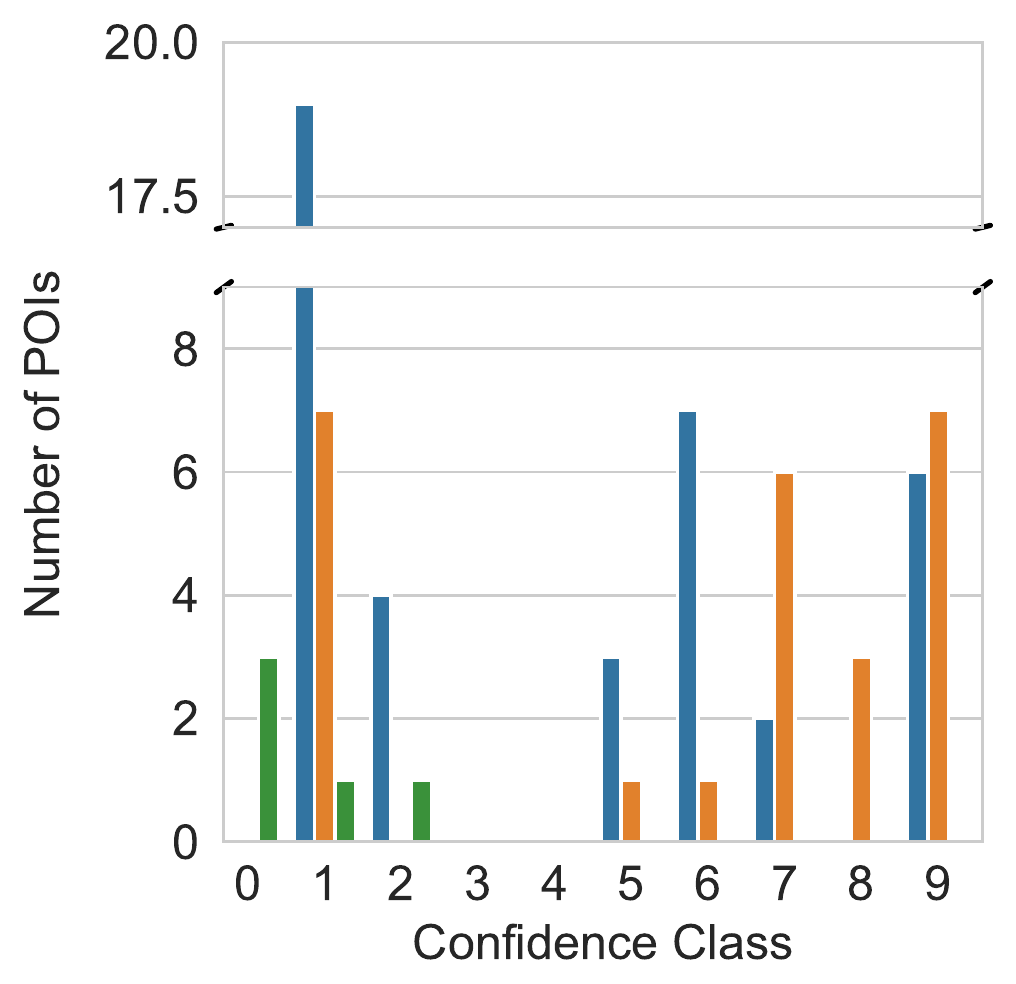}%
    }
    \caption{The distribution of confidence scores of the \acp{POI} identified for each botnet.}
    \label{fig:pois_plot}
\end{figure*}

The distribution of \acp{POI} identified by PinPuppet for each botnet in the evaluation set can be seen in \autoref{fig:pois_plot}. The confidence class in the plot refers to the confidence score of the \acp{POI}. A \ac{POI} with the confidence score $x$ is in the confidence class $c$ when $x\in(\frac{c}{10};\frac{c+1}{10}]$, e.g., for $c=0$  x is in the range $[0.0;0.1]$. Because the confidence score is not used for filtering the port-\acp{POI} (cf. \autoref{sec:pin_puppet_step_poi_candidate_identification}), the plots only contain the data for IP-\acp{POI}.

The first observation is that, as expected, Standalone-\acp{POI}, i.e., the types ``Register (Standalone)'' and ``Memory (Standalone)'', are identified for all botnets. For each botnet, both register and memory \acp{POI} are found with high confidence scores, i.e., a confidence class of eight or nine. 

In contrast, Contiguous-\acp{POI} were only identified for Kelihos and Nugache.
For ZeroAccess and Sality, this is due to them storing IPs internally in a binary format.

Nugache, in contrast, stores its peer list in the registry. For each neighbor, the entry \texttt{HKCU\textbackslash Software\textbackslash GNU\textbackslash Data\textbackslash \{ip\}} is created, where \texttt{\{ip\}} is the decimal ASCII representation of the IP for the respective peer. Thus, it handles the decimal ASCII representation and PinPuppet identifies Contiguous-\acp{POI}.
For Kelihos, we were not able to identify the reason for the Contiguous-\acp{POI} being found. As a sanity check we further investigated the Contiguous-\acp{POI} utilizing our IDA and Ghidra plugins. We were able to determine that two of Kelihos's Contiguous-\acp{POI}, the ones with the lowest confidence score, are part of a \textit{memcpy} function. This makes sense, as a \textit{memcpy} function is a general purpose memory copying function. I.e., it copies values for various unrelated contexts and not only the one we are interested in.

Compared to Standalone-\acp{POI}, only very few Contiguous-\acp{POI} with a high confidence score are found. This is due to the fact that the concept and definition of confidence scores is not as robust for Contiguous-\acp{POI} as it is for Standalone-\acp{POI}. Whereas Standalone-\acp{POI} represent an atomic operation, e.g., a register contains data from the data set or an instruction writes data from the data set, Contiguous-\acp{POI} represent a multi step process. Contiguous-\acp{POI} do reference an instruction using its address, however, patterns are (usually) not satisfied by executing the instruction once. Depending on the implementation and the compiler, other instructions or repeated execution of an instruction might be required to, for example, write an IP address. Both of these problems mean that a definition like with Standalone-\acp{POI} is not possible. Instead, the definition from \autoref{sec:poi_mem_pattern_poi} relies on the number of pattern bytes written which can be skewed by, for example, the malware zeroing memory first. 
All this leads to the fact that the extraction of data from Contiguous-\acp{POI} is limited. The task is basically to extract an unknown pattern from one or more instructions. The challenge here is to determine when an unknown pattern is complete. This can be done for patterns with fixed length or which follow a strict scheme, e.g. ending with a 0 byte. Both is not given for IPs and ports.
Therefore, PinPuppet does not utilize Contiguous-\acp{POI} for extracting peers and instead outputs them for further reverse engineering, the confidence score is less important. For PinPuppet, i.e., peer extraction, the confidence score plays a central part in ensuring that the extracted peers are correct. For reverse engineering, even finding a pattern is useful (cf. \autoref{sec:poi_ensuring_the_quality_of_pois}).

\subsubsection{How well does the confidence score predict the quality of extracted IPs for IP-\acp{POI}?}
\label{sec:evaluation_confidence_score_estimator}
\begin{figure}
    \centering
    \includegraphics[width=0.8\columnwidth]{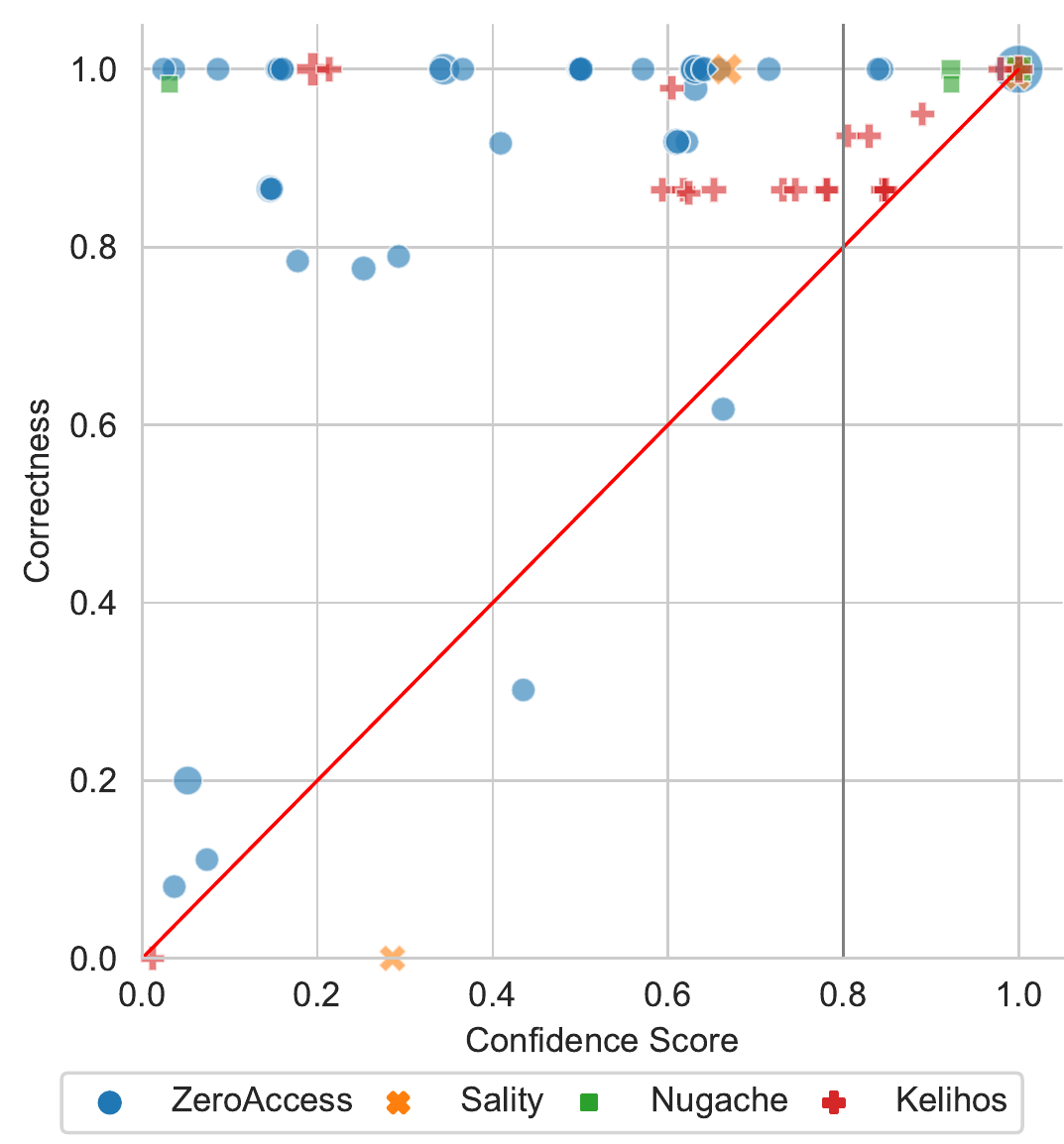}
    \caption{The correctness of each POI.}
    \label{fig:pois_correctness}
\end{figure} 
To better quantify the quality of the results extracted by a \acp{POI}, we introduce the \textit{correctness} metric. For this we are using the previously introduced notation of that $\EP$ is the set of all extracted peers, $\LP$ are all peers in the local botnet, and $\BS$ are all peers on the samples bootstrap list. It is important to note that, due to the local bootstrap peers in our local botnet, $\BS\cap\LP\neq\emptyset$. A peer is seen as correct if and only if it is in the set $\BS\cup\LP$, i.e., all peers that a bot in our local system could know. The correctness for a \ac{POI} is then defined as 

$$\mathit{correctness}=\dfrac{|\EP\cap(\BS\cup\LP)|}{|\EP|}$$

An analysis of the data for all botnets and all \acp{POI} can be seen in \autoref{fig:pois_correctness}. When multiple \acp{POI} of a botnet have the exact same confidence score and $\mathit{correctness}$, they are merged and the size of the corresponding dot increases by one. The red line is the identity line ($x=y$) and the bold, gray, vertical line indicates our confidence score threshold of 0.8. We have determined this confidence score threshold in a parameter study (see \autoref{sec:appendix_confidence_score_threshold} for the results). Our confidence score threshold needs to be below 1.0, as our data sets for searching for \acp{POI} is not complete. As mentioned in \autoref{sec:pin_puppet_step_poi_candidate_identification}, a complete data set would contain all peers from $\BS\cup\SW\cup\PS$, however, we only know $\BS\cup\SW$.

The correctness, as it is defined here, can only be calculated when $\LP$ is known, i.e., with a real-world botnet, the correctness would be unknown. The goal for the confidence score is thus to estimate the correctness and, most importantly, not overestimate it. For example, if a \ac{POI} with a confidence score of 0.9 has a correctness of 0.2, this would mean that a lot of spurious peers were extracted even though the confidence score did not indicate this fact. For \autoref{fig:pois_correctness} this means that the \acp{POI} should be above the red line. As one can see, this is mostly the case. There are only very few \acp{POI} below the red line, however, compared to the total number of \acp{POI} this is negligible. In particular, all \acp{POI} with a confidence score greater than 0.8 (to the right of the green line) are above the red line, i.e., are not overestimated.

Underestimation is expected because our data set $D$ is, as explained in \autoref{sec:pin_puppet_step_poi_candidate_identification}, not necessarily complete. This can be seen in the scatter plot with \acp{POI} being significantly above the red line.

\subsubsection{How effective is PinPuppet in crawling a botnet?}
\input{065_eval_table}

For this part of the evaluation, we are analyzing the types of peers extracted by the crawling primitive. To gather data, the crawling primitive was used to crawl random peers from $\LP$ repeatedly (100 peers in total). We do not use a full crawler implementation to make sure that we only evaluate the shared peer extraction instead of the crawling algorithm, i.e., \ac{DFS}.

To classify the extracted peers, we define three categories: CORRECT, WRONG, and BOOTSTRAP. The first category is $\mathrm{CORRECT}=\EP\cap\LP$. It contains all extracted peers that are in our local botnet and thus correctly extracted by PinPuppet. The second category is $\mathrm{BOOTSTRAP}=\EP\cap(\BS\setminus\LP)$. It contains all extracted peers that are not in our local botnet, but that are still on the bootstrap list. The third category is $\mathrm{WRONG}=\EP\setminus\BS\setminus\LP$. It contains all extracted peers except for the ones that are on the bootstrap list or in our local botnet. Those peers are incorrect, because the botnet could not have known about any of these peers as all bots in our local system know at most the bootstrap list and the local peers.

Number of peers extracted peer category and botnet are given in \autoref{tab:rq3_table}. The results are split up between ``Without confidence score threshold'', i.e., by using all identified \acp{POI}, and ``With confidence score threshold'', i.e., removing results from \acp{POI} with a confidence score below $0.8$. In addition, we show the average confidence score used by the \acp{POI} to extract the results of this category.
For the sake of simplicity, the term "correct" is used in the following if a peer falls into the CORRECT category. This applies analogously to wrong peers and bootstrap peers.

First and most importantly, each crawl cycle extracts correct peers. This is important to note as it shows that the extraction is working. Of course, the more peers that are extracted per crawl cycle, the better a crawler works. However, the maximum number of correct peers that can be extracted per crawl cycle is dependent on the MM-protocol and -mechanism used by the botnet.

Secondly, one can see that applying the confidence score threshold decreases $|\mathrm{WRONG}|$. For ZeroAccess, this results in no wrong peers. For Sality, the number of wrong peers is almost zero. However, sometimes a wrong peer is extracted by a \ac{POI} with a high confidence score. This is not a problem, as it only happens very rarely (in this case three times). For Nugache, exactly one wrong peer remains and looking at the data, the wrong peer is always the same. This can indicate that our bootstrap list is incomplete (which results in this wrong classification) or that this is a value that only appear once, e.g., in an initialization phase. While for Kelihos the confidence score threshold removes a lot of wrong peers, a consistent set of 32 wrong peers remains. Our assumption is that it is a kind of secondary peer list, which is only contacted in certain circumstances. This is supported by the fact that it is always exactly the same 32 peers and 26 of the 32 IPs have already occurred in connection with Kelihos \footnote{\url{https://pastebin.com/9euC4K9N}}, however, we were not able to validate the source.

Looking at the average confidence score, the results are mostly as we have expected. Our expectation was that the average confidence score for $\mathrm{CORRECT}$ and for $\mathrm{BOOTSTRAP}$ would be greater than the one for $\mathrm{WRONG}$. This is mostly the case with the only exception being Sality after applying the confidence score threshold and Kelihos. For Kelihos, we attribute this to the fact that our estimation of the bootstrap list is probably incomplete. For Sality, the reason is the aforementioned peer that is infrequently extracted by a high confidence score \acp{POI}.

Two additional interesting observation are that for Sality and Kelihos, a big part of the bootstrap list is extracted each crawl cycle, and that Kelihos extracts almost all local peers each crawl cycle ($|\LP|=41$ compared to $|\mathrm{CORRECT}|=40.79$). For Sality $|\BS|=740$ but due to the fact that $|\BS\cap\LP|=1$, we get that $|\mathrm{BOOTSTRAP}|=739$ is the maximum amount possible for this metric.

Based on the results outlined above, we can conclude that extracting shared peers from a running malware sample is possible using PinPuppet. Thus, one can use this crawling primitive to develop a fully fledged crawler, for example, utilizing algorithms such as \ac{BFS} or \ac{DFS}~\cite{rossowSoKP2PWNEDModeling2013}, or more advance algorithms such as the \ac{LICA}~\cite{karuppayahAdvancedMonitoringResilient2014}.

%% file: 065_eval_table.tex
\begin{table*}
    \centering
    \scalebox{1}{%
    \begin{tabular}{@{}crrrcrr@{}}
    \toprule
    & & \multicolumn{2}{c}{Without confidence score threshold} & & \multicolumn{2}{c}{With confidence score threshold ($\geq0.8$)} \\
    \cmidrule{3-4} \cmidrule{6-7}
    & & Average Value ($n=100$) & Average confidence score & & Average Value ($n=100$) & Average confidence score \\
    \midrule

\multirow{4}{*}{\rotatebox[origin=c]{90}{ZeroAccess}} & $|\EP|$ & $91.89$ ($\sigma\approx8.62$) & $0.44$ ($n=100$ $\sigma\approx0.01$) & & $17.00$ ($\sigma\approx0.14$) & $0.94$ ($n=100$ $\sigma\approx0.01$) \\
 & $|\mathrm{CORRECT}|$ & $15.91$ ($\sigma\approx0.51$) & $0.48$ ($n=100$ $\sigma\approx0.05$) & & $1.00$ ($\sigma\approx0.14$) & $0.85$ ($n=99$ $\sigma\approx0.01$) \\
 & $|\mathrm{BOOTSTRAP}|$ & $16.00$ ($\sigma\approx0.00$) & $0.62$ ($n=100$ $\sigma\approx0.01$) & & $16.00$ ($\sigma\approx0.00$) & $0.96$ ($n=100$ $\sigma\approx0.01$) \\
 & $|\mathrm{WRONG}|$ & $59.98$ ($\sigma\approx8.39$) & $0.20$ ($n=100$ $\sigma\approx0.01$) & & $0.00$ ($\sigma\approx0.00$) & n/a \\
\midrule

\multirow{4}{*}{\rotatebox[origin=c]{90}{Sality}} & $|\EP|$ & $742.09$ ($\sigma\approx1.11$) & $1.00$ ($n=100$ $\sigma\approx0.00$) & & $741.12$ ($\sigma\approx1.07$) & $1.00$ ($n=100$ $\sigma\approx0.00$) \\
 & $|\mathrm{CORRECT}|$ & $1.94$ ($\sigma\approx0.24$) & $0.83$ ($n=100$ $\sigma\approx0.03$) & & $1.94$ ($\sigma\approx0.24$) & $1.00$ ($n=100$ $\sigma\approx0.00$) \\
 & $|\mathrm{BOOTSTRAP}|$ & $739.00$ ($\sigma\approx0.00$) & $1.00$ ($n=100$ $\sigma\approx0.00$) & & $739.00$ ($\sigma\approx0.00$) & $1.00$ ($n=100$ $\sigma\approx0.00$) \\
 & $|\mathrm{WRONG}|$ & $1.15$ ($\sigma\approx1.05$) & $0.29$ ($n=97$ $\sigma\approx0.02$) & & $0.18$ ($\sigma\approx1.03$) & $1.00$ ($n=3$ $\sigma\approx0.00$) \\
\midrule

\multirow{4}{*}{\rotatebox[origin=c]{90}{Nugache}} & $|\EP|$ & $14.09$ ($\sigma\approx0.72$) & $0.98$ ($n=100$ $\sigma\approx0.00$) & & $13.09$ ($\sigma\approx0.72$) & $0.99$ ($n=100$ $\sigma\approx0.00$) \\
 & $|\mathrm{CORRECT}|$ & $3.45$ ($\sigma\approx0.80$) & $0.98$ ($n=100$ $\sigma\approx0.01$) & & $3.45$ ($\sigma\approx0.80$) & $0.99$ ($n=100$ $\sigma\approx0.00$) \\
 & $|\mathrm{BOOTSTRAP}|$ & $8.64$ ($\sigma\approx0.71$) & $0.99$ ($n=100$ $\sigma\approx0.00$) & & $8.64$ ($\sigma\approx0.71$) & $0.99$ ($n=100$ $\sigma\approx0.00$) \\
 & $|\mathrm{WRONG}|$ & $2.00$ ($\sigma\approx0.00$) & $0.48$ ($n=100$ $\sigma\approx0.00$) & & $1.00$ ($\sigma\approx0.00$) & $0.92$ ($n=100$ $\sigma\approx0.00$) \\
\midrule

\multirow{4}{*}{\rotatebox[origin=c]{90}{Kelihos}} & $|\EP|$ & $301.98$ ($\sigma\approx12.89$) & $0.63$ ($n=100$ $\sigma\approx0.06$) & & $235.79$ ($\sigma\approx1.49$) & $0.85$ ($n=100$ $\sigma\approx0.00$) \\
 & $|\mathrm{CORRECT}|$ & $40.79$ ($\sigma\approx1.49$) & $0.46$ ($n=100$ $\sigma\approx0.12$) & & $40.79$ ($\sigma\approx1.49$) & $0.85$ ($n=100$ $\sigma\approx0.00$) \\
 & $|\mathrm{BOOTSTRAP}|$ & $163.00$ ($\sigma\approx0.00$) & $0.76$ ($n=100$ $\sigma\approx0.03$) & & $163.00$ ($\sigma\approx0.00$) & $0.85$ ($n=100$ $\sigma\approx0.00$) \\
 & $|\mathrm{WRONG}|$ & $98.19$ ($\sigma\approx12.41$) & $0.72$ ($n=100$ $\sigma\approx0.02$) & & $32.00$ ($\sigma\approx0.00$) & $0.85$ ($n=100$ $\sigma\approx0.00$) \\

    \bottomrule
    \end{tabular}%
    }
    \caption{The average number of peers extracted in each crawl cycle by category.}
    \label{tab:rq3_table}
\end{table*}

%% file: 070_relatedwork.tex
\section{Related Work}
\label{sec:rel-work}
In this section, we discuss the related work that are relevant to our work. 
Starting with reverse engineering techniques and tools, up to methods for automated monitoring of botnets.

\paragraph{Reverse-engineering}
IDA and Ghidra are both sophisticated reverse-engineering suites. They give the user many beacons in the code. This goes from basic block representations to the identification of single functions, cross references, resolving of strings, debugger support, and pseudo C code generation. Furthermore, both IDA and Ghidra have a features that automatically identify known functions, even within a binary stripped from symbols and names \footnote{\url{https://www.hex-rays.com/products/ida/tech/flirt/in_depth/}}.
Both tools are designed as manual reverse-engineering tools, but are also platforms to develop custom tools. To the best of our knowledge none if them utilize a \ac{POI} style beacons. However, they would benefit from an implementation of \acp{POI}.

PANDA \cite{pandare}\footnote{\url{https://github.com/panda-re/panda}} is a platform dedicated to dynamic analysis based on QEMU. It stands out because it it allows to record and accurately replay a whole system execution with all running processes. 
This allows the reproducible analysis and execution of samples. 
PANDA can not be used for monitoring a botnet because the sample being analyzed is only communicating with the outside world when recording the system. In particular, PANDA cannot be used to resume after a recording or modify the execution of an existing recording\footnote{\url{https://github.com/panda-re/panda/blob/master/panda/docs/manual.md}}.

Taint analysis is used to mark certain data to trace the data flow in a binary from data sources, like user input, to sinks where the input is used, e.g., in SQL statements.
Taint analysis can, for example, be used to determine where a user input is being processed. This way, possible SQL injections and XSS can be detected automatically \cite{taintSql}.
The main difference to the \acp{POI} is that taint analysis marks all instructions that interact with (parts of) the data from one origin (source).
These markers can be a kind of beacon for the reverse-engineer~\cite{bincat}\footnote{\url{https://github.com/airbus-seclab/bincat}}. 
The \ac{POI} approach in contrast finds and scores individual instructions that access the data $d \in D$. Furthermore, no sources need to be defined as with taint analysis, but only the relevant data.

\paragraph{Automated P2P Botnet Monitoring}

A honeypot may be an alternative for collecting peers of a botnet. 
The honeypot runs the malware sample and records the IPs of outgoing and incoming messages.
As the peer running in the honeypot is integrated into the botnet's overlay network, botnet peers start communicating with the local peer. This serves as a kind of passive peer enumeration, however, compared to active crawling, this method is inherently slower and less directed.

%% file: 999_appendix.tex
\newpage
\begin{appendix}

\begin{table*}
    \centering
    \begin{tabular}{rcc}
        \toprule
        Botnet & SHA-256 & MD5 \\
        \midrule
        ZeroAccess & \texttt{69e966e730557fde8fd84317cdef1ece00a8bb3470c0b58f3231e170168af169} & \texttt{ea039a854d20d7734c5add48f1a51c34} \\
        Sality & \texttt{0b283b3ee065c2a1a5d9b5fef691be7b70cf5c5f1371f5a6653ec35a998602a0} & \texttt{d35cf3c2335666ac0be74f93c5f5172f} \\
        Nugache & \texttt{fcb69486e3f4745d90a26446ec07925eb2f1f8812a3a676495d72cd1a9951f68} & \texttt{0c859cfad2fa154f007042a1dca8d75b} \\
        Kelihos & \texttt{b71a4a57d21742797ec9079c745e2f884cb9379717069189bf0839078b0e2c62} & \texttt{9b68b45afa269ba1b0c01749fa4b942f}\\
        \bottomrule
    \end{tabular}
    \caption{Hashes of the samples used.}
    \label{tab:sample_hashes}
\end{table*}

\section{Hashes of malware Samples used for the Evaluation}
\label{sec:appendix_hashes}
\autoref{tab:sample_hashes} contains the SHA256 and MD-5 hashes of the samples used in our evaluation.

\section{Confidence Score Threshold Parameter Study}
\label{sec:appendix_confidence_score_threshold}
\begin{figure}[b]
\centering
    \includegraphics[width=\columnwidth]{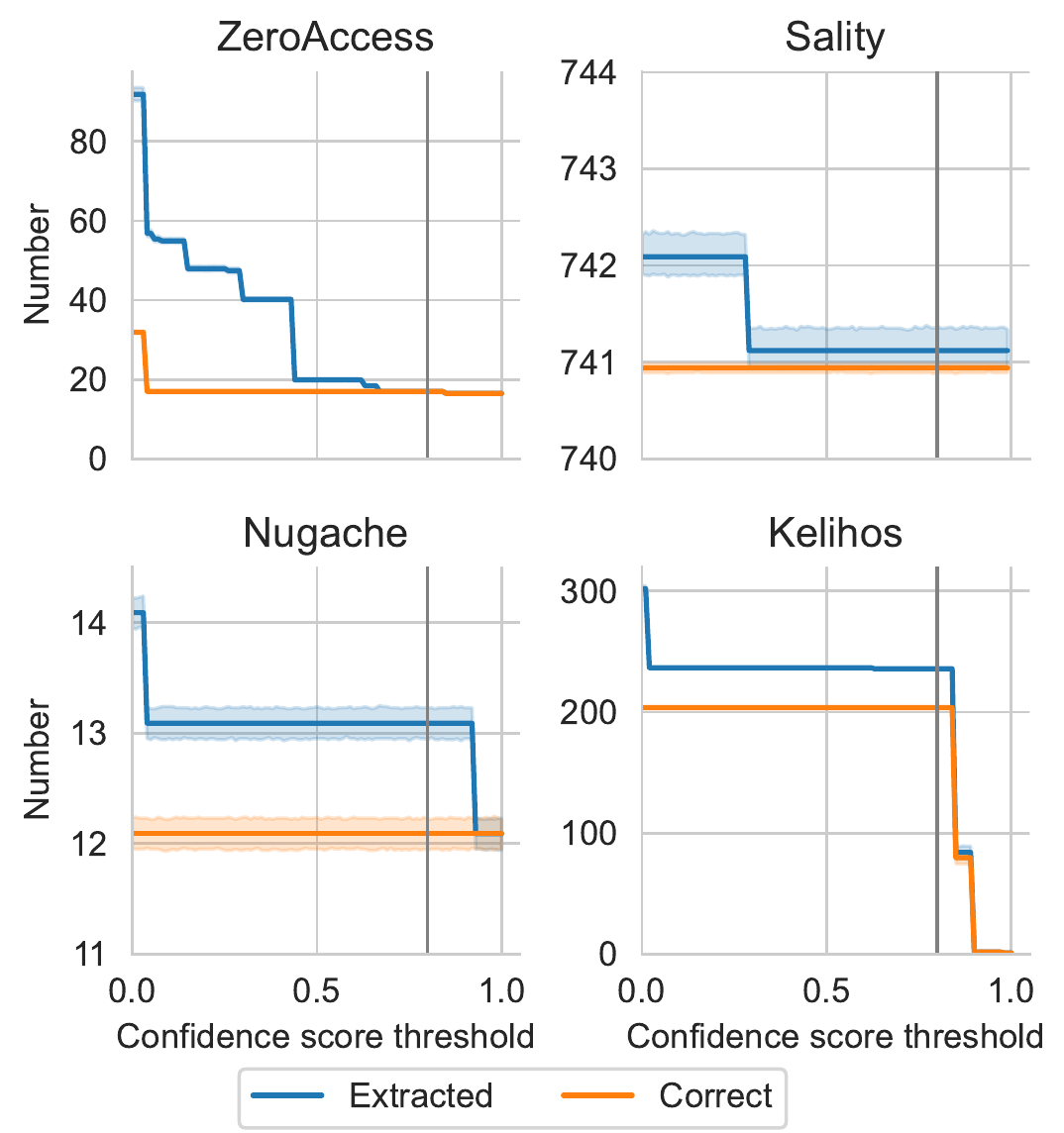}
    \caption{Peer extraction analysis}
    \label{fig:threshold_analysis2}
\end{figure}

The confidence score threshold needs to be carefully chosen to find the balance between a) wrong peers being extracted and b) how many peers are extracted. We have analyzed these metrics in \autoref{fig:threshold_analysis2} and have determined 0.8 as the threshold (marked with a gray line). As can be seen in the plot, $0.8$ strikes a good balance between the number of wrong extracted peers, i.e., the distance between the upper and the lower lines, and the total number of extracted peers. $0.9$, for example, would not have been suitable as then, Kelihos only extracts very few peers.

\section{Overhead Measurement Results}
\label{sec:appendix_overhead}
\autoref{tab:overhead} contains our raw measurement data of the overhead introduced by using Pin to trace a sample. For each botnet, the measurement was repeated $5$ times. 
The evaluation in \autoref{sec:eval_pinpuppet_speed} is based on that data.
    
\input{table_overhead_short}

\begin{figure*}[h]
    \centering
    \includegraphics[width=1\textwidth]{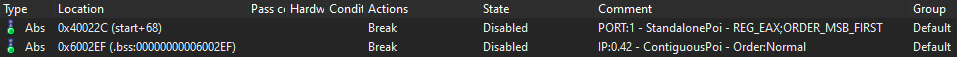}
    \includegraphics[width=1\textwidth]{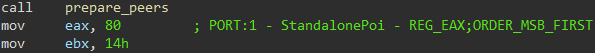}
    \caption{IDA plugin overview.}
    \label{fig:ida_plugin}
\end{figure*}
    
\begin{figure*}[h]
    \centering
    \includegraphics[width=1\textwidth]{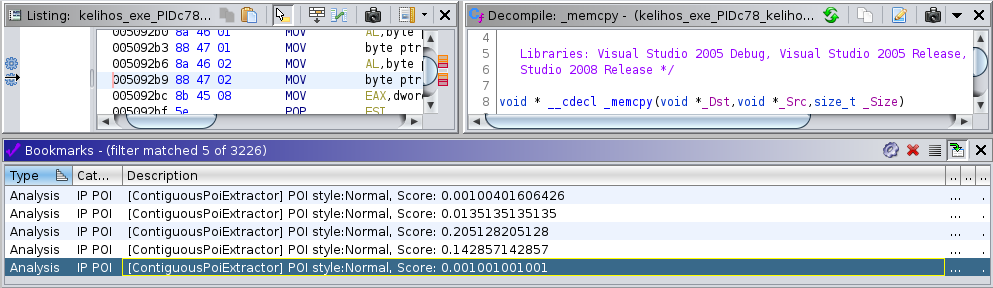}
    \caption{Ghidra plugin overview.}
    \label{fig:ghidra_plugin}
\end{figure*}

\section{IDA and Ghidra Plugins for Importing and Displaying POIs}
\label{sec:appendix_plugins}
It is important that the image loaded in IDA is aligned with the one from which the POIs have been extracted.
The IDA plugin is a python script which is read by IDA. Via the menu file->Script File..., this script can be selected.
Afterwards a dialog box appears which allows the selection of a \ac{JSON} file where the exported POIs are stored.
These \acp{POI} are then imported into IDA as disabled breakpoints. They are then visible as an overview in the breakpoint view and also in the disassembler view (see \autoref{fig:ida_plugin}).

For Ghidra, the plugin is also implemented in Python. After enabling the Plugin in the script manager, it can be activated in the toolbar. Like with IDA, one is presented with a dialog box where one needs to select the \acp{POI} \ac{JSON} file. Afterwards, the \acp{POI} show up as bookmarks in the disassembled code (see \autoref{fig:ghidra_plugin}).

\end{appendix}

%% file: table_overhead_short.tex
\begin{table*}
    \centering
    \begin{tabular}{@{}rrrrcrrr@{}}
        \toprule
        & \multicolumn{3}{c}{Without puppeteering} & & \multicolumn{3}{c}{With puppeteering} \\
        \cmidrule{2-4} \cmidrule{5-8}
        & $T_0$ & $T_1-T_0$ & $T_2-T_1$ & & $T_0$ & $T_1-T_0$ & $T_2-T_1$ \\
        \midrule
        
ZeroAccess  & $12.21$ ($\sigma\approx0.24$) & $20.01$ ($\sigma\approx0.00$) & $20.00$ ($\sigma\approx0.00$) & & $14.81$ ($\sigma\approx0.03$) & $19.97$ ($\sigma\approx0.01$) & $20.00$ ($\sigma\approx0.00$)\\

Sality  & $335.58$ ($\sigma\approx0.56$) & $32.57$ ($\sigma\approx0.00$) & $43.71$ ($\sigma\approx0.00$) & & $689.36$ ($\sigma\approx82.87$) & $32.51$ ($\sigma\approx0.10$) & $43.44$ ($\sigma\approx0.15$)\\

Nugache  & $0.61$ ($\sigma\approx0.32$) & $405.60$ ($\sigma\approx0.00$) & $405.60$ ($\sigma\approx0.00$) & & $58.67$ ($\sigma\approx32.80$) & $406.75$ ($\sigma\approx0.08$) & $406.27$ ($\sigma\approx0.09$)\\

Kelihos  & $0.78$ ($\sigma\approx0.02$) & $200.30$ ($\sigma\approx0.00$) & $600.29$ ($\sigma\approx0.00$) & & $0.81$ ($\sigma\approx0.02$) & $213.25$ ($\sigma\approx0.31$) & $601.19$ ($\sigma\approx0.03$)\\

        \bottomrule
    \end{tabular}

    \caption{The overhead measurements for each botnet ($n=5$).}
    \label{tab:overhead}
\end{table*}

%% file: 000_Puppeteering.bbl

\begin{thebibliography}{28}


\ifx \showCODEN    \undefined \def \showCODEN     #1{\unskip}     \fi
\ifx \showDOI      \undefined \def \showDOI       #1{#1}\fi
\ifx \showISBNx    \undefined \def \showISBNx     #1{\unskip}     \fi
\ifx \showISBNxiii \undefined \def \showISBNxiii  #1{\unskip}     \fi
\ifx \showISSN     \undefined \def \showISSN      #1{\unskip}     \fi
\ifx \showLCCN     \undefined \def \showLCCN      #1{\unskip}     \fi
\ifx \shownote     \undefined \def \shownote      #1{#1}          \fi
\ifx \showarticletitle \undefined \def \showarticletitle #1{#1}   \fi
\ifx \showURL      \undefined \def \showURL       {\relax}        \fi
\providecommand\bibfield[2]{#2}
\providecommand\bibinfo[2]{#2}
\providecommand\natexlab[1]{#1}
\providecommand\showeprint[2][]{arXiv:#2}

\bibitem[\protect\citeauthoryear{Babil, Mehani, Boreli, and Kaafar}{Babil
  et~al\mbox{.}}{2013}]%
        {antiTaint13}
\bibfield{author}{\bibinfo{person}{Golam~Sarwar Babil},
  \bibinfo{person}{Olivier Mehani}, \bibinfo{person}{Roksana Boreli}, {and}
  \bibinfo{person}{Mohamed-Ali Kaafar}.} \bibinfo{year}{2013}\natexlab{}.
\newblock \showarticletitle{On the effectiveness of dynamic taint analysis for
  protecting against private information leaks on Android-based devices}. In
  \bibinfo{booktitle}{\emph{2013 International Conference on Security and
  Cryptography (SECRYPT)}}. \bibinfo{pages}{1--8}.
\newblock


\bibitem[\protect\citeauthoryear{Banescu, Collberg, Ganesh, Newsham, and
  Pretschner}{Banescu et~al\mbox{.}}{2016}]%
        {symbolicExecObfuscation}
\bibfield{author}{\bibinfo{person}{Sebastian Banescu},
  \bibinfo{person}{Christian Collberg}, \bibinfo{person}{Vijay Ganesh},
  \bibinfo{person}{Zack Newsham}, {and} \bibinfo{person}{Alexander
  Pretschner}.} \bibinfo{year}{2016}\natexlab{}.
\newblock \showarticletitle{Code obfuscation against symbolic execution
  attacks}. In \bibinfo{booktitle}{\emph{Proceedings of the 32nd Annual
  Conference on Computer Security Applications}}. \bibinfo{pages}{189--200}.
\newblock


\bibitem[\protect\citeauthoryear{Biondi, Rigo, Zennou, and Mehrenberger}{Biondi
  et~al\mbox{.}}{2017}]%
        {bincat}
\bibfield{author}{\bibinfo{person}{Philippe Biondi},
  \bibinfo{person}{Rapha{\"e}l Rigo}, \bibinfo{person}{Sarah Zennou}, {and}
  \bibinfo{person}{Xavier Mehrenberger}.} \bibinfo{year}{2017}\natexlab{}.
\newblock \showarticletitle{BinCAT: purrfecting binary static analysis}. In
  \bibinfo{booktitle}{\emph{Symposium sur la s{\'e}curit{\'e} des technologies
  de l’information et des communications}}.
\newblock


\bibitem[\protect\citeauthoryear{Bruening, Zhao, and Amarasinghe}{Bruening
  et~al\mbox{.}}{2012}]%
        {dynamorio}
\bibfield{author}{\bibinfo{person}{Derek Bruening}, \bibinfo{person}{Qin Zhao},
  {and} \bibinfo{person}{Saman Amarasinghe}.} \bibinfo{year}{2012}\natexlab{}.
\newblock \showarticletitle{Transparent dynamic instrumentation}. In
  \bibinfo{booktitle}{\emph{Proceedings of the 8th ACM SIGPLAN/SIGOPS
  conference on Virtual Execution Environments}}. \bibinfo{pages}{133--144}.
\newblock


\bibitem[\protect\citeauthoryear{Calvet, Davis, Fernandez, Marion, St-Onge,
  Guizani, Bureau, and Somayaji}{Calvet et~al\mbox{.}}{2010}]%
        {Calvet2010}
\bibfield{author}{\bibinfo{person}{Joan Calvet}, \bibinfo{person}{Carlton~R.
  Davis}, \bibinfo{person}{Jose~M. Fernandez}, \bibinfo{person}{Jean-Yves
  Marion}, \bibinfo{person}{Pier-Luc St-Onge}, \bibinfo{person}{Wadie Guizani},
  \bibinfo{person}{Pierre-Marc Bureau}, {and} \bibinfo{person}{Anil Somayaji}.}
  \bibinfo{year}{2010}\natexlab{}.
\newblock \showarticletitle{{The case for in-the-lab botnet experimentation:
  creating and taking down a 3000-node botnet}}. In
  \bibinfo{booktitle}{\emph{Proceedings of the 26th Annual Computer Security
  Applications Conference}}. \bibinfo{pages}{141--150}.
\newblock
\showISBNx{9781450301336}


\bibitem[\protect\citeauthoryear{Cavallaro, Saxena, and Sekar}{Cavallaro
  et~al\mbox{.}}{2007}]%
        {antiTaint07}
\bibfield{author}{\bibinfo{person}{Lorenzo Cavallaro}, \bibinfo{person}{Prateek
  Saxena}, {and} \bibinfo{person}{R Sekar}.} \bibinfo{year}{2007}\natexlab{}.
\newblock \showarticletitle{Anti-taint-analysis: Practical evasion techniques
  against information flow based malware defense}.
\newblock \bibinfo{journal}{\emph{Secure Systems Lab at Stony Brook University,
  Tech. Rep}} (\bibinfo{year}{2007}), \bibinfo{pages}{1--18}.
\newblock


\bibitem[\protect\citeauthoryear{Chen, Andersen, Mao, Bailey, and Nazario}{Chen
  et~al\mbox{.}}{2008}]%
        {antiDebugging}
\bibfield{author}{\bibinfo{person}{Xu Chen}, \bibinfo{person}{Jon Andersen},
  \bibinfo{person}{Z~Morley Mao}, \bibinfo{person}{Michael Bailey}, {and}
  \bibinfo{person}{Jose Nazario}.} \bibinfo{year}{2008}\natexlab{}.
\newblock \showarticletitle{Towards an understanding of anti-virtualization and
  anti-debugging behavior in modern malware}. In \bibinfo{booktitle}{\emph{2008
  IEEE international conference on dependable systems and networks with FTCS
  and DCC (DSN)}}. IEEE, \bibinfo{pages}{177--186}.
\newblock


\bibitem[\protect\citeauthoryear{Clause, Li, and Orso}{Clause
  et~al\mbox{.}}{2007}]%
        {dynamictaintDytan}
\bibfield{author}{\bibinfo{person}{James Clause}, \bibinfo{person}{Wanchun Li},
  {and} \bibinfo{person}{Alessandro Orso}.} \bibinfo{year}{2007}\natexlab{}.
\newblock \showarticletitle{Dytan: a generic dynamic taint analysis framework}.
  In \bibinfo{booktitle}{\emph{Proceedings of the 2007 international symposium
  on Software testing and analysis}}. \bibinfo{pages}{196--206}.
\newblock


\bibitem[\protect\citeauthoryear{D'Elia, Coppa, Nicchi, Palmaro, and
  Cavallaro}{D'Elia et~al\mbox{.}}{2019}]%
        {deliaSoKUsingDynamic2019}
\bibfield{author}{\bibinfo{person}{Daniele~Cono D'Elia},
  \bibinfo{person}{Emilio Coppa}, \bibinfo{person}{Simone Nicchi},
  \bibinfo{person}{Federico Palmaro}, {and} \bibinfo{person}{Lorenzo
  Cavallaro}.} \bibinfo{year}{2019}\natexlab{}.
\newblock \showarticletitle{{{SoK}}: {{Using Dynamic Binary Instrumentation}}
  for {{Security}} ({{And How You May Get Caught Red Handed}})}. In
  \bibinfo{booktitle}{\emph{Proceedings of the 2019 {{ACM Asia Conference}} on
  {{Computer}} and {{Communications Security}}}} \emph{(\bibinfo{series}{Asia
  {{CCS}} '19})}. \bibinfo{publisher}{{Association for Computing Machinery}},
  \bibinfo{address}{{New York, NY, USA}}, \bibinfo{pages}{15--27}.
\newblock
\showISBNx{978-1-4503-6752-3}


\bibitem[\protect\citeauthoryear{Dolan-Gavitt, Hodosh, Hulin, Leek, and
  Whelan}{Dolan-Gavitt et~al\mbox{.}}{2015}]%
        {pandare}
\bibfield{author}{\bibinfo{person}{Brendan Dolan-Gavitt}, \bibinfo{person}{Josh
  Hodosh}, \bibinfo{person}{Patrick Hulin}, \bibinfo{person}{Tim Leek}, {and}
  \bibinfo{person}{Ryan Whelan}.} \bibinfo{year}{2015}\natexlab{}.
\newblock \showarticletitle{Repeatable reverse engineering with PANDA}. In
  \bibinfo{booktitle}{\emph{Proceedings of the 5th Program Protection and
  Reverse Engineering Workshop}}. \bibinfo{pages}{1--11}.
\newblock


\bibitem[\protect\citeauthoryear{Eilam}{Eilam}{2011}]%
        {eilam2011reversing}
\bibfield{author}{\bibinfo{person}{Eldad Eilam}.}
  \bibinfo{year}{2011}\natexlab{}.
\newblock \bibinfo{booktitle}{\emph{Reversing: secrets of reverse
  engineering}}.
\newblock \bibinfo{publisher}{John Wiley \& Sons}.
\newblock


\bibitem[\protect\citeauthoryear{Gorgovan}{Gorgovan}{2016}]%
        {gorgovanEscapingDynamoRIOPin2021}
\bibfield{author}{\bibinfo{person}{Cosmin Gorgovan}.}
  \bibinfo{year}{2016}\natexlab{}.
\newblock \bibinfo{title}{Escaping {{DynamoRIO}} and {{Pin}}}.
\newblock
\newblock
\urldef\tempurl%
\url{https://github.com/lgeek/dynamorio\_pin\_escape}
\showURL{%
Retrieved May 6, 2021 from \tempurl}


\bibitem[\protect\citeauthoryear{Karuppayah}{Karuppayah}{2018}]%
        {karuppayah2018advanced}
\bibfield{author}{\bibinfo{person}{Shankar Karuppayah}.}
  \bibinfo{year}{2018}\natexlab{}.
\newblock \bibinfo{booktitle}{\emph{Advanced Monitoring in P2P Botnets: A Dual
  Perspective}}.
\newblock \bibinfo{publisher}{Springer}.
\newblock
\showISBNx{9789811090493}


\bibitem[\protect\citeauthoryear{Karuppayah, Fischer, Rossow, and
  M{\"u}hlh{\"a}user}{Karuppayah et~al\mbox{.}}{2014}]%
        {karuppayahAdvancedMonitoringResilient2014}
\bibfield{author}{\bibinfo{person}{Shankar Karuppayah},
  \bibinfo{person}{Mathias Fischer}, \bibinfo{person}{Christian Rossow}, {and}
  \bibinfo{person}{Max M{\"u}hlh{\"a}user}.} \bibinfo{year}{2014}\natexlab{}.
\newblock \showarticletitle{On Advanced Monitoring in Resilient and
  Unstructured {{P2P}} Botnets}. In \bibinfo{booktitle}{\emph{2014 {{IEEE
  International Conference}} on {{Communications}} ({{ICC}})}}.
  \bibinfo{pages}{871--877}.
\newblock
\showISSN{1938-1883}


\bibitem[\protect\citeauthoryear{Kerkers, Santanna, and Sperotto}{Kerkers
  et~al\mbox{.}}{2014}]%
        {kelihos}
\bibfield{author}{\bibinfo{person}{Max Kerkers}, \bibinfo{person}{Jos{\'e}~Jair
  Santanna}, {and} \bibinfo{person}{Anna Sperotto}.}
  \bibinfo{year}{2014}\natexlab{}.
\newblock \showarticletitle{Characterisation of the kelihos. b botnet}. In
  \bibinfo{booktitle}{\emph{IFIP International Conference on Autonomous
  Infrastructure, Management and Security}}. Springer, \bibinfo{pages}{79--91}.
\newblock


\bibitem[\protect\citeauthoryear{Kirsch, Zhechev, Bierbaumer, and
  Kittel}{Kirsch et~al\mbox{.}}{2018}]%
        {kirschPwINPwningIntel2018a}
\bibfield{author}{\bibinfo{person}{Julian Kirsch}, \bibinfo{person}{Zhechko
  Zhechev}, \bibinfo{person}{Bruno Bierbaumer}, {and} \bibinfo{person}{Thomas
  Kittel}.} \bibinfo{year}{2018}\natexlab{}.
\newblock \showarticletitle{{{PwIN}} \textendash{} {{Pwning Intel piN}}: {{Why
  DBI}} Is {{Unsuitable}} for {{Security Applications}}}. In
  \bibinfo{booktitle}{\emph{Computer {{Security}}}}
  \emph{(\bibinfo{series}{Lecture {{Notes}} in {{Computer Science}}})},
  \bibfield{editor}{\bibinfo{person}{Javier Lopez}, \bibinfo{person}{Jianying
  Zhou}, {and} \bibinfo{person}{Miguel Soriano}} (Eds.).
  \bibinfo{publisher}{{Springer International Publishing}},
  \bibinfo{address}{{Cham}}, \bibinfo{pages}{363--382}.
\newblock


\bibitem[\protect\citeauthoryear{Lin, Zhang, and Xu}{Lin et~al\mbox{.}}{2010}]%
        {autoRe}
\bibfield{author}{\bibinfo{person}{Zhiqiang Lin}, \bibinfo{person}{Xiangyu
  Zhang}, {and} \bibinfo{person}{Dongyan Xu}.} \bibinfo{year}{2010}\natexlab{}.
\newblock \showarticletitle{Automatic reverse engineering of data structures
  from binary execution}. In \bibinfo{booktitle}{\emph{Proceedings of the 11th
  Annual Information Security Symposium}}.
\newblock


\bibitem[\protect\citeauthoryear{Luk, Cohn, Muth, Patil, Klauser, Lowney,
  Wallace, Reddi, and Hazelwood}{Luk et~al\mbox{.}}{2005}]%
        {intelpin}
\bibfield{author}{\bibinfo{person}{Chi-Keung Luk}, \bibinfo{person}{Robert
  Cohn}, \bibinfo{person}{Robert Muth}, \bibinfo{person}{Harish Patil},
  \bibinfo{person}{Artur Klauser}, \bibinfo{person}{Geoff Lowney},
  \bibinfo{person}{Steven Wallace}, \bibinfo{person}{Vijay~Janapa Reddi}, {and}
  \bibinfo{person}{Kim Hazelwood}.} \bibinfo{year}{2005}\natexlab{}.
\newblock \showarticletitle{Pin: building customized program analysis tools
  with dynamic instrumentation}.
\newblock \bibinfo{journal}{\emph{Acm sigplan notices}} \bibinfo{volume}{40},
  \bibinfo{number}{6} (\bibinfo{year}{2005}), \bibinfo{pages}{190--200}.
\newblock


\bibitem[\protect\citeauthoryear{Neville and Gibb}{Neville and Gibb}{2013}]%
        {nevilleZeroAccessIndepth2013}
\bibfield{author}{\bibinfo{person}{Alan Neville} {and} \bibinfo{person}{Ross
  Gibb}.} \bibinfo{year}{2013}\natexlab{}.
\newblock \showarticletitle{{{ZeroAccess Indepth}}}.
\newblock \bibinfo{journal}{\emph{Symantec Security Response}}
  (\bibinfo{year}{2013}).
\newblock


\bibitem[\protect\citeauthoryear{Newsome, Brumley, Franklin, and Song}{Newsome
  et~al\mbox{.}}{2006}]%
        {replayer}
\bibfield{author}{\bibinfo{person}{James Newsome}, \bibinfo{person}{David
  Brumley}, \bibinfo{person}{Jason Franklin}, {and} \bibinfo{person}{Dawn
  Song}.} \bibinfo{year}{2006}\natexlab{}.
\newblock \showarticletitle{Replayer: Automatic protocol replay by binary
  analysis}. In \bibinfo{booktitle}{\emph{Proceedings of the 13th ACM
  conference on Computer and communications security}}.
  \bibinfo{pages}{311--321}.
\newblock


\bibitem[\protect\citeauthoryear{Nguyen-Tuong, Guarnieri, Greene, Shirley, and
  Evans}{Nguyen-Tuong et~al\mbox{.}}{2005}]%
        {taintSql}
\bibfield{author}{\bibinfo{person}{Anh Nguyen-Tuong},
  \bibinfo{person}{Salvatore Guarnieri}, \bibinfo{person}{Doug Greene},
  \bibinfo{person}{Jeff Shirley}, {and} \bibinfo{person}{David Evans}.}
  \bibinfo{year}{2005}\natexlab{}.
\newblock \showarticletitle{Automatically hardening web applications using
  precise tainting}. In \bibinfo{booktitle}{\emph{IFIP International
  Information Security Conference}}. Springer, \bibinfo{pages}{295--307}.
\newblock


\bibitem[\protect\citeauthoryear{Or-Meir, Nissim, Elovici, and Rokach}{Or-Meir
  et~al\mbox{.}}{2019}]%
        {surveyMalware}
\bibfield{author}{\bibinfo{person}{Ori Or-Meir}, \bibinfo{person}{Nir Nissim},
  \bibinfo{person}{Yuval Elovici}, {and} \bibinfo{person}{Lior Rokach}.}
  \bibinfo{year}{2019}\natexlab{}.
\newblock \showarticletitle{Dynamic Malware Analysis in the Modern Era—A
  State of the Art Survey}.
\newblock \bibinfo{journal}{\emph{ACM Computing Surveys (CSUR)}}
  \bibinfo{volume}{52}, \bibinfo{number}{5}, Article \bibinfo{articleno}{88}
  (\bibinfo{date}{Sept.} \bibinfo{year}{2019}), \bibinfo{numpages}{48}~pages.
\newblock
\showISSN{0360-0300}


\bibitem[\protect\citeauthoryear{Rossow, Andriesse, Werner, {Stone-Gross},
  Plohmann, Dietrich, and Bos}{Rossow et~al\mbox{.}}{2013}]%
        {rossowSoKP2PWNEDModeling2013}
\bibfield{author}{\bibinfo{person}{Christian Rossow}, \bibinfo{person}{Dennis
  Andriesse}, \bibinfo{person}{Tillmann Werner}, \bibinfo{person}{Brett
  {Stone-Gross}}, \bibinfo{person}{Daniel Plohmann},
  \bibinfo{person}{Christian~J. Dietrich}, {and} \bibinfo{person}{Herbert
  Bos}.} \bibinfo{year}{2013}\natexlab{}.
\newblock \showarticletitle{{{SoK}}: {{P2PWNED}} - {{Modeling}} and
  {{Evaluating}} the {{Resilience}} of {{Peer}}-to-{{Peer Botnets}}}. In
  \bibinfo{booktitle}{\emph{2013 {{IEEE Symposium}} on {{Security}} and
  {{Privacy}}}}. \bibinfo{pages}{97--111}.
\newblock
\showISSN{1081-6011}


\bibitem[\protect\citeauthoryear{Stover, Dittrich, Hernandez, and
  Dietrich}{Stover et~al\mbox{.}}{2007}]%
        {stoverAnalysisStormNugache2007}
\bibfield{author}{\bibinfo{person}{S. Stover}, \bibinfo{person}{D. Dittrich},
  \bibinfo{person}{John Hernandez}, {and} \bibinfo{person}{S. Dietrich}.}
  \bibinfo{year}{2007}\natexlab{}.
\newblock \showarticletitle{Analysis of the Storm and Nugache Trojans: {{P2P}}
  Is Here}.
\newblock \bibinfo{journal}{\emph{login Usenix Mag.}}  \bibinfo{volume}{32}
  (\bibinfo{year}{2007}).
\newblock


\bibitem[\protect\citeauthoryear{{Symantec}}{{Symantec}}{2019}]%
        {Symantec2019}
\bibfield{author}{\bibinfo{person}{{Symantec}}.}
  \bibinfo{year}{2019}\natexlab{}.
\newblock \bibinfo{title}{Internet Security Threat Report 2019}.
\newblock
  \bibinfo{howpublished}{\url{https://docs.broadcom.com/doc/istr-24-2019-en}}.
\newblock


\bibitem[\protect\citeauthoryear{Votipka, Rabin, Micinski, Foster, and
  Mazurek}{Votipka et~al\mbox{.}}{2019}]%
        {votipkaObservationalInvestigationReverse2019a}
\bibfield{author}{\bibinfo{person}{Daniel Votipka}, \bibinfo{person}{Seth~M.
  Rabin}, \bibinfo{person}{Kristopher Micinski}, \bibinfo{person}{Jeffrey~S.
  Foster}, {and} \bibinfo{person}{Michelle~L. Mazurek}.}
  \bibinfo{year}{2019}\natexlab{}.
\newblock \showarticletitle{An {{Observational Investigation}} of {{Reverse
  Engineers}}' {{Processes}}}.
\newblock \bibinfo{journal}{\emph{arXiv:1912.00317 [cs]}} (\bibinfo{date}{Nov.}
  \bibinfo{year}{2019}).
\newblock


\bibitem[\protect\citeauthoryear{Yan, Zhang, and Ansari}{Yan
  et~al\mbox{.}}{2008}]%
        {yan2008revealing}
\bibfield{author}{\bibinfo{person}{Wei Yan}, \bibinfo{person}{Zheng Zhang},
  {and} \bibinfo{person}{Nirwan Ansari}.} \bibinfo{year}{2008}\natexlab{}.
\newblock \showarticletitle{Revealing Packed Malware}.
\newblock \bibinfo{journal}{\emph{IEEE Security Privacy}} \bibinfo{volume}{6},
  \bibinfo{number}{5} (\bibinfo{year}{2008}), \bibinfo{pages}{65--69}.
\newblock


\bibitem[\protect\citeauthoryear{You and Yim}{You and Yim}{2010}]%
        {malwareObfuscation}
\bibfield{author}{\bibinfo{person}{Ilsun You} {and} \bibinfo{person}{Kangbin
  Yim}.} \bibinfo{year}{2010}\natexlab{}.
\newblock \showarticletitle{Malware Obfuscation Techniques: A Brief Survey}.
\newblock \bibinfo{journal}{\emph{Proceedings - 2010 International Conference
  on Broadband, Wireless Computing Communication and Applications, BWCCA
  2010}}, \bibinfo{pages}{297--300}.
\newblock


\end{thebibliography}
